\documentclass[traditabstract]{aa} 
\bibpunct{(}{)}{;}{a}{}{,}

\newcommand{\ergps}{erg\thinspace s$^{-1}$}
\newcommand{\phpspsqcm}{ph\thinspace cm$^{-2}$\thinspace s$^{-1}$}
\newcommand{\ergpspsqcm}{erg\thinspace s$^{-1}$\thinspace cm$^{-2}$}
\newcommand{\ergpspsqcmpkeV}{erg\thinspace s$^{-1}$\thinspace cm$^{-2}$\thinspace keV$^{-1}$}
\newcommand{\psqcm}{cm$^{-2}$}
\newcommand{\nH}{$N_{\rm H}$}

\usepackage[varg]{txfonts}
\usepackage{graphicx}
\usepackage{subfig}
\begin{document}
\title{Origin of the diffuse 4-8 keV emission in M82}


\author{K. Iwasawa\inst{1,2}
  \and
C. Norman\inst{3,4}
\and
R. Gilli\inst{5}
\and
P. Gandhi\inst{6}
\and
M. A. Per\'ez-Torres\inst{7,8,9}
}

\institute{Institut de Ci\`encies del Cosmos (ICCUB), Universitat de Barcelona (IEEC-UB), Mart\'i i Franqu\`es, 1, 08028 Barcelona, Spain
         \and
         ICREA, Pg. Llu\'is Companys 23, 08010 Barcelona, Spain
         \and
         Department of Physics and Astronomy, Johns Hopkins University, Baltimore, MD 21218, USA
         \and
         Space Telescope Science Institute, 3700 San Martin Drive, Baltimore, MD 21218, USA
         \and
         INAF - Osservatorio di Astrofisica e Scienza dello Spazio di Bologna, Via P. Gobetti 93/3, 40129 Bologna, Italy
         \and
         School of Physics and Astronomy, University of Southampton, Highfield, Southampton SO17\,1BJ, UK
         \and
         Instituto de Astrof\'isica de Andaluc\'ia (CSIC), Glorieta de la Astronom\'ia s/n, 18008, Granada, Spain
         \and
         Facultad de Ciencias, Universidad de Zaragoza, Pedro Cerbuna 12, E-50009 Zaragoza, Spain
         \and
         School of Sciences, European University Cyprus, Diogenes street, Engomi, 1516 Nicosia, Cyprus
}


 
\abstract{We present the first spatially resolved, X-ray spectroscopic study of the 4-8 keV diffuse emission found in the central part of the nearby starburst galaxy M82 on a few arcsecond scales. The new details that we see allow a number of important conclusions to be drawn on the nature of the hot gas and its origin as well as feedback on the interstellar medium. We use archival data from the {\it Chandra X-ray Observatory} with an exposure time of 570 ks. The Fe\,{\sc xxv} emission at 6.7 keV, expected from metal-enriched hot gas, is enhanced only in a limited area close to the starburst disc and is weak or almost absent over the rest of the diffuse emission, resulting in spatial variations in equivalent width from $<0.1$ keV to 1.9 keV. This shows the presence of non-thermal emission due to inverse Compton scattering of the far-infrared photons by radio emitting cosmic ray electrons. The morphological resemblance between the diffuse X-ray, radio, and far-infrared emission maps support this concept. Our decomposition of the diffuse emission spectrum indicates that $\sim $70\% of the 4-8 keV luminosity originates from the inverse Compton emission. The metal-rich hot gas with a temperature of $\simeq 5$ keV makes a minor contribution to the 4-8 keV continuum, but it accounts for the majority of the observed Fe\,{\sc xxv} line.
This hot gas appears to emerge from the circumnuclear starburst ring and fill the galactic chimneys identified as mid-infrared and radio emission voids. The energetics argument suggests that much of the supernova energy in the starburst site has gone into creating of the chimneys and is transported to the halo. We argue that a hot, rarefied environment produced by strong supernova feedback results in displacing the brightest X-ray and radio supernova remnants which are instead found to reside in giant molecular clouds.
We find a faint X-ray source with a radio counterpart, close to the kinematic centre of the galaxy and we carefully examine the possibility that this source is a low-luminosity active galactic nucleus in an advection-dominated accretion flow phase. 
}  

\titlerunning{Diffuse Fe K emission in M82}
\authorrunning{K. Iwasawa}
\maketitle
%

\section{Introduction}

A fundamental ingredient of the superwind model of \citet{chevalier85}  is a hot fluid with a temperature of $\sim 10^8$ K, created by an injection of kinetic energy mainly from supernovae, and to a lesser degree from stellar winds of massive stars, propagating into a low-density medium in a starburst region. To maintain the kinetic energy for driving a galactic-scale wind, this hot fluid has to be minimally radiative with a long cooling time.  The superwind  could be observed in hard X-rays above a few keV. The nearby starburst galaxy M82 shows a large-scale galactic wind at various wavelengths and is an ideal target to examine the wind-driving mechanism. Earlier X-ray imaging of M82 at the arcsecond resolution with the {\it Chandra X-ray Observatory} ({\it Chandra}) resolved out bright X-ray binaries which dominate the hard X-ray band and revealed that there is residual diffuse emission surrounding the starburst disc \citep{griffiths00}. The spectrum of the diffuse gas exhibits a high-ionisation Fe K line from Fe\,{\sc xxv}, characteristic of the thermal emission of a temperature of a few keV, and thus has been identified with the hot fluid as envisaged in the superwind model. \citet{strickland09} argued that the observed Fe\,{\sc xxv} luminosity agrees with the expectation of the \citet{chevalier85} model. However, if the hard X-ray diffuse emission spectrum is to be explained by thermal emission, the Fe\,{\sc xxv} line appears too weak, requiring low Fe metallicity of 0.2-0.3 $Z_\sun$ \citep{strickland07,liu14}, in contrast with a highly enriched interstellar medium enriched by core-collapse supernovae in starburst galaxies such as M82. Recent findings of several X-ray supernova remnants (SNRs), which also emit strong Fe\,{\sc xxv} but are not resolved in the previous study \citep{iwasawa21}, reduce the Fe\,{\sc xxv} luminosity originating from the diffuse emission further. A possible explanation is that the continuum of the Fe\,{\sc xxv}-emitting gas is much fainter and the majority of the observed continuum emission comes from an extra source such as non-thermal emission \citep[e.g.][]{strickland09}. Although non-thermal emission due to inverse Compton scattering of far-infrared (FIR) photons by cosmic ray electrons have long been proposed for the hard X-ray emission in M82 \citep{hargrave74,schaaf89,moran97}, the presence of bright X-ray binaries hindered a good measurement of the true diffuse emission properties before {\it Chandra}. Even with the {\it Chandra} imaging, the inverse Compton emission was disfavoured by \citet{strickland07} for being too faint. Given the $\gamma $-ray detection from M82 \citep{abdo10} and the importance of cosmic-ray driven winds in starburst galaxies, the starburst-driven non-thermal process warrants a serious consideration.

In this work, we present an updated spatially resolved spectral analysis of the X-ray diffuse emission in the 4-8 keV band using the archival {\it Chandra} data. The Fe\,{\sc xxv} emission at 6.7 keV is of primary interest, but the cold Fe K emission at 6.4 keV is also present in the diffuse emission spectrum \citep{strickland07,liu14}. We examine the spatial distributions of these Fe lines and discuss their implications for the origin of the 4-8 keV emission, which seems to support the presence of both the inverse Compton emission and also the metal-rich hot gas. 

The distance to M82 is assumed to be 3.6 Mpc \citep{freedman94}. The angular scale is thus 17.4 pc arcsec$^{-1}$ (1 kpc corresponds to approximately 1 arcmin).

\section{Observations}

\begin{table}
\caption{{\it Chandra} observations of M82.}
\label{tab:obslog}
\centering
\begin{tabular}{cccccc}
  \hline\hline
Date & ObsID & ACIS & Exp. & $\theta$ & $\phi$ \\
  \hline
1999-09-20 & 361 & I3 & 33 & 0.3 & 26 \\
1999-09-20 & 1302 & I3 & 15 & 0.3 & 26\\
  \hline
2002-06-18 & 2933 & S3 & 18 & 0.6 & 1 \\
2009-06-24 & 10542 & S3 & 118 & 1.4 & 223 \\
2009-07-01 & 10543 & S3 & 118 & 1.1 & 179 \\
2009-07-07 & 10925 & S3 & 45 & 2.6 & 21 \\
2009-07-08 & 10544 & S3 & 74 & 2.6 & 21 \\
2010-06-17 & 11104 & S3 & 10 & 0.4 & 50 \\
2010-07-20 & 11800 & S3 & 17 & 2.7 & 1 \\
2010-07-23 & 10545 & S3 & 95 & 2.7 & 8 \\
2012-08-09 & 13796 & S3 & 20 & 0.3 & 169 \\
2014-02-03 & 16580 & S3 & 47 & 1.2 & 323 \\
2015-01-20 & 16023 & S3 & 10 & 1.2 & 356 \\
\hline
\end{tabular}
\tablefoot{Date: the date when each observation started. ACIS: the ACIS CCD chip on which the central part of the M82 was placed. Exp.: useful exposure time in units of $10^3$ seconds. $\theta $, $\phi $: Off-axis angle in arcminutes and roll angle in degrees of the central position of Fig. 1 in each observation. The first two ACIS-I observations were included only for the imaging analysis.}
\end{table}

We used the same 13 {\it Chandra} ACIS observations of M82 as in \citet{iwasawa21} where details of the observations can be found. These observations were selected from all the {\it Chandra} ACIS observations of M82 for relatively small off axis angles (0.3-2.7 arcsec) -for the central starburst region- so that the point spread function (PSF) has small distortion. The major part of this work relating to spectral analysis makes use of the 11 observations with the ACIS-S3, which provide a consistent spectral resolution. The two early AICS-I observations in 1999 were used only for the imaging analysis. The total exposure time of the 11 ACIS-S3 observations is 570 ks and the two ACIS-I observations add an extra 52 ks.

{\it Chandra} data reduction and image analysis were carried out by CIAO 4.12 \citep{fruscione06} using the calibration files in CALDB~4.53. Spectral analysis was performed using HEASoft XSPEC \citep{blackburn99,arnaud96}.

\section{Results}

\subsection{Image of the 4-8 keV diffuse emission}

\begin{figure*}
\centerline{\includegraphics[width=0.85\textwidth,angle=0]{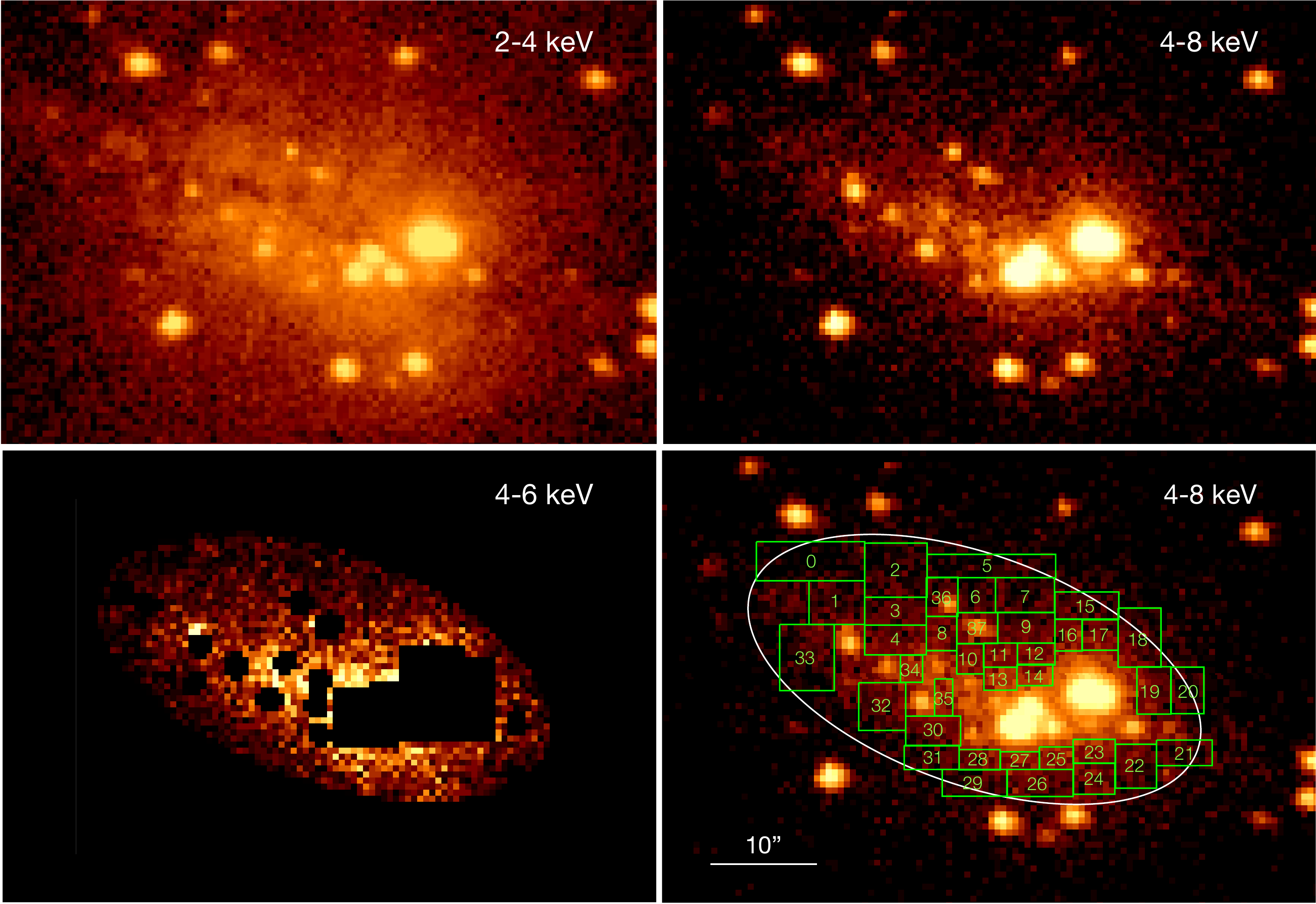}}
\caption{{\it Chandra} X-ray images of the central part of M82 integrated over the 13 observations. Top-left: 2-4 keV image; Top-right: 4-8 keV band; The colour scale of these two images is in logarithmic with the brightest point sources are saturated. Bottom-left: The 4-6 keV image of the ellipse where the data of the diffuse emission spectrum were taken. The detected discrete sources were masked. The region where the brightest ULXs along with a few others are clustered is masked in a block. This image is in a linear scale as the dynamic range of the brightness is narrow: 1-19, as a result of the removal of discrete sources. Bottom-right: The same image as Top right overlaid by the ellipse and the 38 region segments for spatially resolved spectral study.}
\label{fig:e24e48img}
\end{figure*}

The 4-8 keV {\it Chandra} image integrated over all the 13 observations is shown in Fig. 1. The diffuse emission, as well as a number of discrete sources, is visible. It lies along the starburst disc where the discrete sources are concentrated. Its extension along the major axis (PA $\simeq 70^{\circ}$) is  $\sim 700$ pc, coinciding with the disc size. The height of the extension above and below the disc is less than 200 pc and the overall shape is elliptical. We adopt an ellipse of semi-major and semi-minor axis lengths of $19\farcs 4 \times 9\farcs 4$ (=338 pc $\times $ 164 pc) with the major axis inclined to $PA=70^{\circ}$, which includes most of the 4-8 keV diffuse emission.

The spectral data were taken from this ellipse, masking the individual discrete sources. We refer to the 21 sources detected within the ellipse in \citet{iwasawa21}, Xp0 through Xp20. The western part where the bright ultraluminous X-ray sources (ULXs) are clustered was masked in a block so that a sufficiently large area around them is masked to avoid their emission in the PSF wings as well as the bright core emission. This block contains Xp12 through Xp17. These masks for the discrete sources take up 23\% of the entire ellipse area of 591 arcsec$^{2}$. The 4-6 keV continuum image from the ellipse with the masking is also shown in Fig. 1. We note that the 4-6 keV image of the diffuse emission alone has a relatively flat brightness distribution, in the absence of peaky discrete sources, and is shown in a linear scale.

Compared with the 4-8 keV emission, the diffuse emission at lower energies rises sharply towards the disc plane and is more extended, as illustrated by the 2-4 keV image. This soft X-ray extended emission is considered to be the entrained, shocked interstellar medium (ISM), forming part of the large-scale soft X-ray winds. The central part, near the starburst disc in particular, is obscured and subject to suppression by absorption at the lower energies. Even in the 2-4 keV image, the light suppression is visible along the disc. This absorption effect of the order of \nH\ of a few times of $10^{22}$ \psqcm\ practically diminishes above 4 keV and all the discrete sources embedded in the starburst disc are visible in the 4-8 keV image. Likewise, the diffuse emission above 4 keV is largely unaffected by absorption, except for a small part with excess obscuration.

\subsection{Total diffuse emission spectrum}

\subsubsection{Thermal emission lines}

\begin{figure}
  \centerline{\includegraphics[width=0.4\textwidth,angle=0]{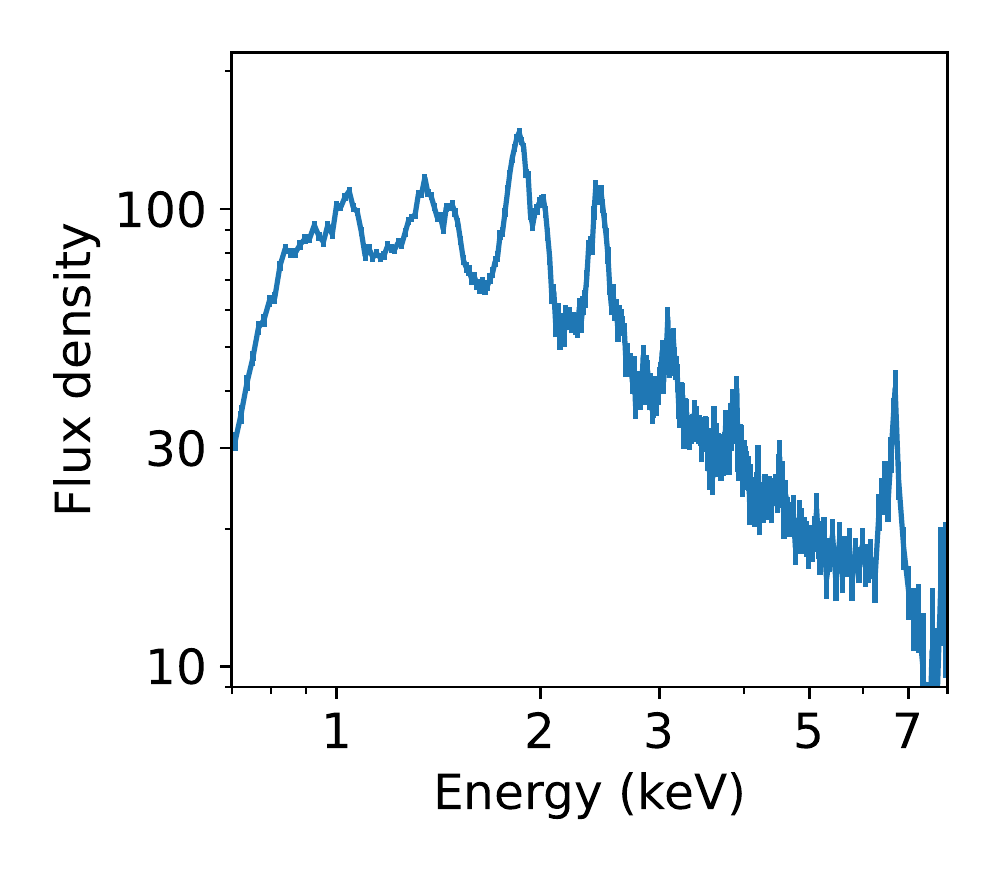}}
\vspace{-5mm}  \centerline{\includegraphics[width=0.35\textwidth,angle=0]{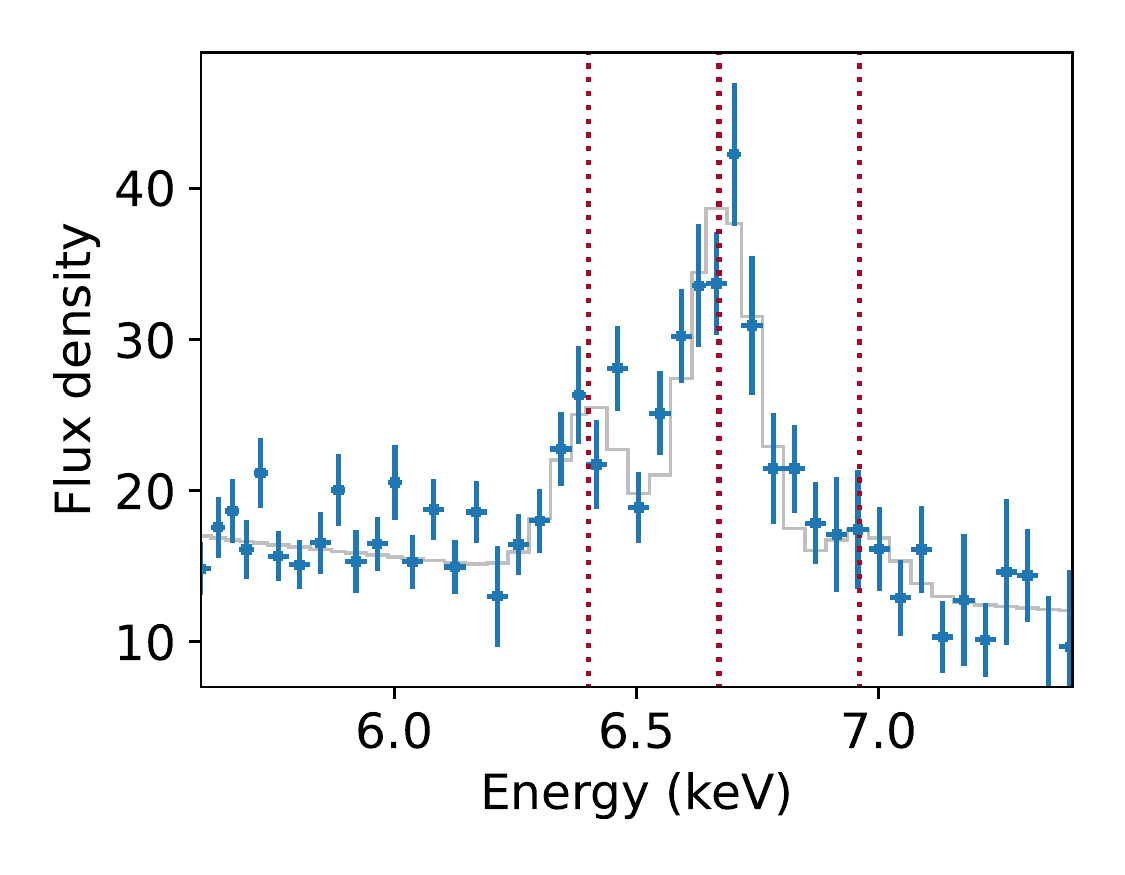}}
    \caption{Diffuse emission spectrum of the central part of M82, obtained from the {\it Chandra} ACIS-S. The flux density is in units of $10^{-14}$ \ergpspsqcmpkeV. Upper panel: The 0.7-8 keV spectrum; Lower panel: The Fe K band spectrum of the same data. The three vertical dotted-lines indicates the expected centroid energies of 6.4 keV, 6.7 keV and 6.96 keV for cold Fe K, Fe\,{\sc xxv} and Fe\,{\sc xxvi}, respectively. The grey line indicates the fitted model of a power-law and three Gaussian lines.}
\label{fig:fnbb}
\end{figure}

Although the soft X-ray emission spectrum is beyond the scope of this work, it is worth illustrating the contrasting characteristics of the emission-line spectrum above and below 4 keV. The broad-band spectrum (Fig. \ref{fig:fnbb}) shows strong He-like and H-like K-shell emission from Ne, Mg, Si, S, Ar and Ca in the 1-4 keV range. At energies above 4 keV, the most prominent spectral feature is the Fe K complex in 6-7 keV (Fig. \ref{fig:fnbb}), dominated by Fe\,{\sc xxv} at $\sim 6.7$ keV. As previously noted by \citet{strickland07,liu14}, a cold Fe K line at 6.4 keV is also present. A line-like excess is detected marginally at 4.5 keV, although the origin is unclear. Since its detection is marginal and is not essential for this work, we leave further details in Appendix B. For all the spectral fits presented below unless stated otherwise, a spectrum was binned so that each bin contains at least one count. A spectral fit was performed on background-corrected data by comparing with a model folded through the detector response, using the C-statistic \citep{cash79}. 

\begin{figure}
\centerline{\includegraphics[width=0.37\textwidth,angle=0]{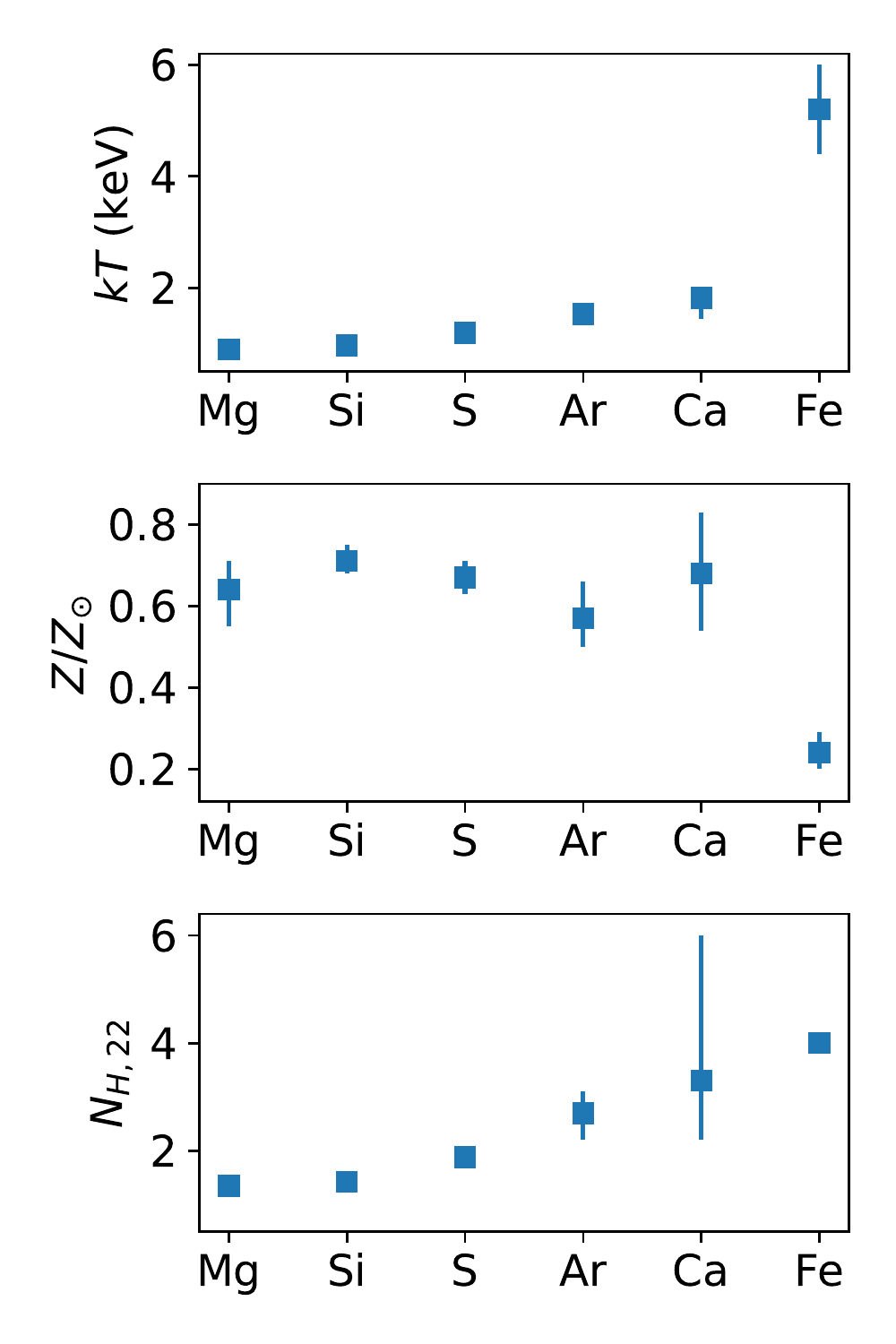}}
\caption{Temperature ($kT$), metallicity ($Z$) and absorbing column density (\nH ) obtained from fitting the {\tt apec} to the individual line spectra. The \nH\ for the Fe K band spectrum was assumed to be $4\times 10^{22}$ \psqcm, approximately the middle value obtained for the Ca spectrum, as the energy range of the data is insensitive to \nH $< 10^{23}$ \psqcm. }
\label{fig:spfit}
\end{figure}

We examined temperature and metallicity inferred from the atomic lines of each element by fitting the thermal emission spectrum of {\tt apec}\footnote{{\tt apec} is the code to calculate a thermal emission spectrum \citep{foster12}. Here, gas in a collisional ionisation equilibrium is assumed.} in a narrow energy range, as follows: 1.1-1.7 keV for Mg; 1.5-2.4 keV for Si; 2.1-3 keV for S; 2.7-3.5 keV for Ar; and 3.3-4.3 keV for Ca. With the restricted energy ranges, a temperature and metallicity were inferred by each element independently. A He-like and H-like line ratio mainly drives the temperature determination for Mg, Si, S, for which the emission lines dominate the continuum. This is less true for Ar and Ca, since the slope of strong continuum emission affects the temperature determination. We did not look into the Ne lines, as the energy range of this line also contains substantial emission from the Fe L complex around 1 keV. Cold absorption was fitted. The results are illustrated in Fig. \ref{fig:spfit}. The metallicity remains compatible between the elements at $\simeq 0.6\, Z_{\odot}$. The gas temperatures inferred from individual elements remain in a limited range of $kT \simeq 0.9$-1.8 keV with a moderately increasing trend as the selected energy range moves higher. The absorption also shows the same increasing trend. Both indicate spectral hardening towards higher energies. This is likely a manifestation of an increasing contribution of a hard continuum component rather than a true rise of temperature or absorption. Therefore we consider that the temperature and \nH\ for Ar and Ca lines, which lie above the strong continuum, are likely similar to those inferred for S emission. 

In contrast, the Fe K line spectrum gives a much higher temperature and a lower metallicity. Fitting to the 6-7.2 keV spectrum gives $kT=4.6$ (3.8-5.7) keV and $0.21^{+0.6}_{-0.4} Z_{\odot}$. In the fit, an additional narrow Gaussian line at 6.4 keV is included. As the energy range is not sensitive to absorption smaller than \nH $\sim 10^{23}$ \psqcm, the central \nH\ value obtained for the Ca K band was assumed but it has no effect on the result. This abrupt change in temperature and metallicity above and below 4 keV, in addition to the morphological difference in image, signals a transition between the two diffuse emission components of different origins taking place at around 4 keV. The increasing $kT$ and \nH\ towards the transition energy suggest spectral hardening which is likely caused by the harder component and/or more absorption. The soft X-ray wind emission therefore lies in the foreground of the harder component exhibiting the Fe K emission.


\subsubsection{Fe K complex}

The Fe K line complex (Fig. \ref{fig:fnbb}) is described here by multiple Gaussians. The dominant Fe\,{\sc xxv} is most likely of thermal origin. It is a triplet that is not resolved at the CCD resolution. Fitting a single Gaussian to a simulated {\tt apec} spectrum gives a centroid energy of 6.68 keV with a slight broadening of $\sigma\sim 30$ eV, when $kT$ is in the range of 4-6 keV (Appendix A), which is inferred from the {\tt apec} fit to the Fe K band. The observed data show the centroid energy of $6.67\pm 0.01$ keV and the width of $\sigma = 58\pm 17$ eV, slightly broader than expected. Two other Gaussian lines for a cold Fe K line at 6.4 keV and Fe\,{\sc xxvi} at 6.96 keV were added, assuming they are narrow and unresolved.
Respective line intensities were fitted and the results are shown in Table \ref{tab:feklines}. The continuum is modelled by a power-law by fitting the 4.1-8 keV data (the lower energy bound at 4.1 keV was chosen to eliminate the Ca emission). The absorbing 
 \nH\  of  $4\times 10^{22}$ \psqcm\ and the marginally detected 4.5 keV line are also included. The power-law slope, energy index\footnote{We use energy index, $\alpha $, for a spectral slope appropriate for a spectrum in flux density unit, as defined as $f_{\rm E}\propto E^{-\alpha}$, where $f_{\rm E}$ is flux density and $E$ is energy. The conventional photon index, $\Gamma $, is related as $\Gamma = 1+\alpha $.}, was found to be $\alpha = 1.28\pm 0.12$.

Notwithstanding the weak detection of Fe~{\sc xxvi}, the Fe~{\sc xxvi}/Fe~{\sc xxv} ratio of $0.14\pm 0.09$ indicates a temperature of $kT\approx 4.5\pm 1$ keV (Appendix A), in agreement with that inferred from the {\tt apec} fit. However, the Fe\,{\sc xxv} line broadening as observed above may indicate a contribution from a lower temperature emission, as discussed later, and thus the true temperature of the gas may be higher than that inferred from the apparent Fe\,{\sc xxv}/Fe\,{\sc xxvi} ratio (see further details in Sect. 4).


\begin{table}
\caption{Fe K line complex of the diffuse emission spectrum.}
\label{tab:feklines}
\centering
\begin{tabular}{ccccc}
  \hline\hline
  Energy & Intensity & {\sl EW} & $L_{\rm line}$ & ID \\
  \hline
  6.4 & $1.8\pm 0.3$ & $0.13\pm 0.02$  & 2.9 & Cold Fe \\
  6.67 & $4.9\pm 0.5$ & $0.38\pm 0.04$ & 8.1 & Fe\,{\sc xxv} \\
  6.96 & $0.68 \pm 0.43$ & $0.06\pm 0.04$ & 1.1 & Fe\,{\sc xxvi} \\
  \hline
\end{tabular}
\tablefoot{Line energies, width ($\sigma $) and EW are in units of keV. Line intensities are in units of $10^{-6}$ ph~cm$^{-2}$~s$^{-1}$. Line luminosities are in units of $10^{37}$ \ergps. They are as observed from the spectrum from the region masking the discrete source apertures.}
\end{table}

\subsection{Spectral variations across the diffuse emission}

\subsubsection{Region segments}

Here, we investigate spectral variations across the 4-8 keV diffuse emission of M82. We set 38 rectangular region-segments covering the diffuse emission (Fig. \ref{fig:e24e48img}) and examined the individual spectra taken there. These region segments were placed by avoiding the discrete sources, except for the two transient sources (the reason for this exception is described below). We label them as dX0 through dX37. The segments dX0 through dX20 are on the northern side of the starburst disc while dX21 through dX33 on the southern side. The two segments, dX34 and dX35, lie within the starburst disc, placed in the small gaps between crowded discrete sources.

The two additional segments, dX36 and dX37, contain transient sources, Xp7 and Xp11 as labelled in \citet{iwasawa21}, respectively. Since the transient sources were in quiescence, no excess emission is visible at their positions (Appendix C) and we used the data for their quiescent state. As the flaring of Xp7 occurred in the two early ACIS-I observations used only for imaging, the exposure time of the dX36 spectrum remains the same as the spectra of the other segments, integrated over all the 11 ACIS-S observations. As for Xp11, the transient flaring occurred in the three observations in 2010. The spectrum of dX37 was then produced, omitting those observations, and its resulted exposure time is 450 ks, as opposed to the full 570 ks.

The segment sizes were chosen to minimise the dispersion of source counts contained between the individual segments where allowed by the positions of the discrete sources. As a result, the 4-8 keV counts of those segments range between 150 and 400 counts, with the mean of 274 counts and the standard deviation of 72 counts. The broad-band spectra of these 38 region segments are shown in Appendix F (Fig. \ref{fig:dxbbsp}).

\subsubsection{Mapping the spectral variations}

\begin{figure*}
\centerline{\includegraphics[width=0.95\textwidth,angle=0]{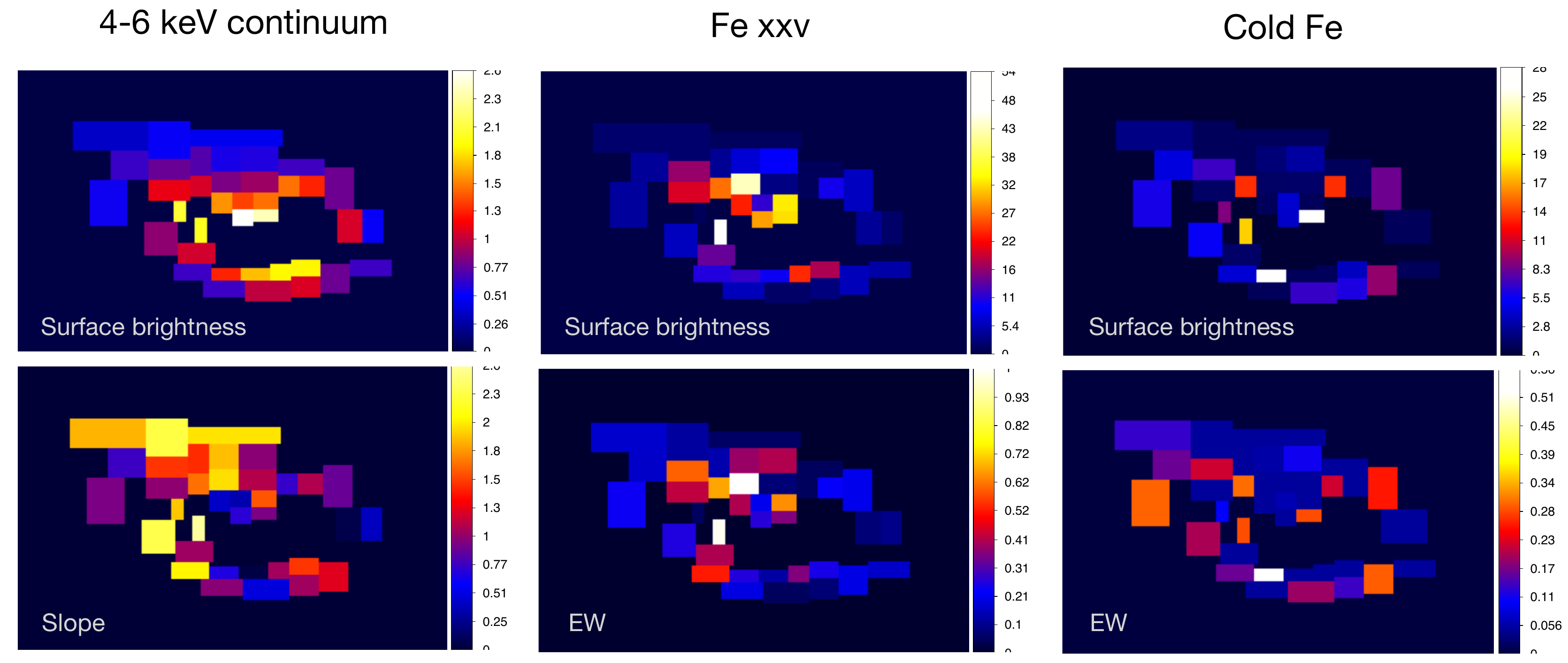}}
\caption{Spatial variations of spectral properties over the hard-band diffuse emission. Left: Surface brightness (upper panel) in units of $10^{-15}$ \ergpspsqcm\thinspace arcsec$^{-2}$ and energy index of power-law continuum (lower panel) in the 4-6 keV band; Middle: Surface brightness of 6.67 keV Fe~{\sc xxv} emission in units of $10^{-9}$ \phpspsqcm\thinspace arcsec$^{-2}$ and line EW in units of keV; and Right: Surface brightness and line EW of 6.4 keV cold Fe K emission in the same units as the Fe\,{\sc xxv} emission.}
\label{fig:maps}
\end{figure*}

\begin{figure}
\centerline{\includegraphics[width=0.5\textwidth,angle=0]{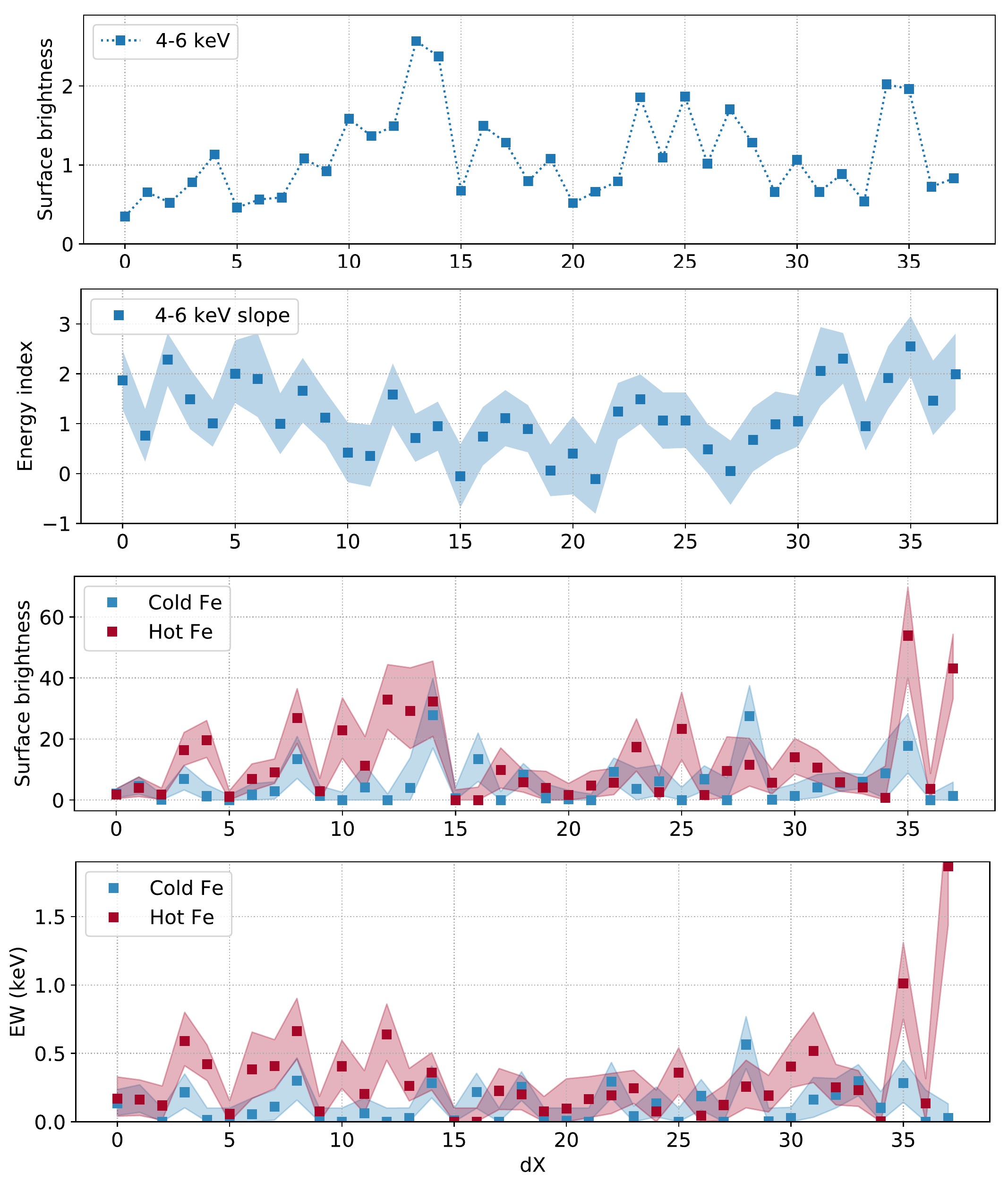}}
\caption{Values of the respective region segments in each map in Fig. \ref{fig:maps} are plotted. From the top to the bottom, surface brightness of the 4-6 keV continuum in units of $10^{-15}$ \ergpspsqcm\thinspace arcsec$^{-2}$; the 4-6 keV energy index, $\alpha $; surface brightness of Cold Fe (6.4 keV, in blue) and Hot Fe (Fe\,{\sc xxv} at 6.67 keV, in red) in units of $10^{-9}$ ph\,s$^{-1}$\,cm$^{-2}$\,arcsec$^{-2}$; and EW of Cold Fe (6.4 keV) and Hot Fe (6.67 keV) in units of keV.}
\label{fig:map_values}
\end{figure}

\begin{figure}
\centerline{\includegraphics[width=0.5\textwidth,angle=0]{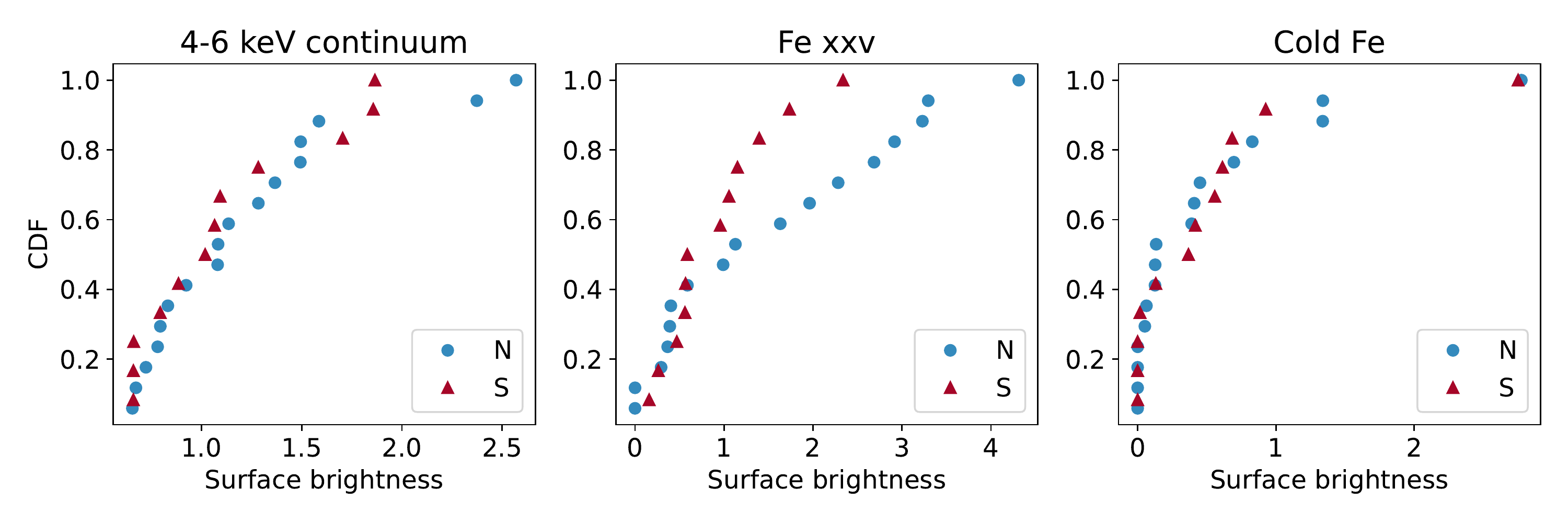}}
\caption{Separate CDFs of the surface brightness of the 4-6 keV continuum, Fe\,{\sc xxv} and cold Fe (from left to right) in the northern (N, blue circles) and southern (S, red triangles) parts with respect to the starburst disc plane. These plots were made, using 29 segments with the 4-6 keV surface brightness cut (see text). The 4-6 keV continuum surface brightness is in units of $10^{-15}$ erg\,s$^{-1}$\,cm$^{-2}$\,arcsec$^{-2}$ while the Fe line surface brightness is in units of $10^{-8}$ ph\,s$^{-1}$\,cm$^{-2}$\,arcsec$^{-2}$}
\label{fig:ecdf_surbri}
\end{figure}

Our primary interest is to investigate spatial variations of the Fe K lines. We modelled the 4-8 keV spectrum of each segment with a power-law continuum and two Gaussian lines at 6.4 keV and 6.67 keV. No absorption is included. The Fe lines were assumed to be narrow at those energies and their normalisations were fitted. The continuum flux of each region was estimated in the 4-6 keV band from the fit. The power-law energy index, $\alpha $, measured for the continuum in the same energy range was also recorded for each region.

Based on the above fits, we constructed surface brightness maps of the 4-6 keV continuum, cold Fe and Fe\,{\sc xxv} emission over the 38 region segments, as shown in Fig. \ref{fig:maps}. The maps of 4-6 keV continuum slope and equivalent width (EW) of the two Fe lines are also shown. The same values given in the maps with the 68\% error intervals are plotted in the order of the segment number in Fig. \ref{fig:map_values}. A few spectra with a 4-6 keV slope close to zero indicate that they might be more absorbed than the rest since they have \nH\ of $\sim 10^{23}$ \psqcm\ (Appendix D). 

The continuum emission shows high surface brightness near the starburst disc which declines gradually towards larger radii, in agreement with the 4-6 keV count image shown in Fig. \ref{fig:e24e48img}. With respect to the disc plane, the brightness distribution appears to be similar between the northern and southern sides, when discounting the lowest surface brightness regions on the north-east corner. On the contrary, the Fe\,{\sc xxv} emission is markedly stronger on the northern side, particularly evident in its eastern part. Such a north-south asymmetry is not obvious in the cold Fe emission.

The above observation regarding the north-south brightness symmetry and asymmetry is also illustrated by a comparison of cumulative distribution functions (CDFs) of surface brightness of those quantities (Fig. \ref{fig:ecdf_surbri}). The diffuse continuum emission has the lowest surface brightness at the north-eastern outskirt (see Fig. \ref{fig:maps}). On applying the surface brightness cut of $>0.06\times 10^{-15}$ \ergpspsqcm\thinspace arcsec$^{-2}$, the area of the faintest emission is discarded, leaving 17/12 segments (out of 22/14 segments) on the northern/southern sides. The segments dX34 and dX35 located in the disc plane were not considered. The median values of the brightness of the northern and southern segments are found to be comparable: $1.08\times 10^{-9}$ \phpspsqcm\thinspace arcsec$^{-2}$ and $1.04\times 10^{-9}$ \phpspsqcm\thinspace arcsec$^{-2}$, respectively. Their CDFs of the 4-6 keV surface brightness are indeed similar, except for the brightest end.The north-south brightness distributions in the continuum for the selected segments are balanced. It can then be used as a reference to test any north-south asymmetry in the Fe line emission when the same selected segments are used. The CDFs for Fe\,{\sc xxv} clearly diverge between the northern and southern segments towards higher brightness, verifying the brighter northern side.

A line detection at the confidence level of $>95$\% is found to occur when line intensity exceeds $1\times 10^{-7}$ \phpspsqcm. There are 15 segments showing this relatively strong Fe\,{\sc xxv} detection and most of them are located on the northern side (Fig. \ref{fig:map_values}) and connected to each other (Fig. \ref{fig:maps}), suggesting a continuous distribution. Unmatched continuum and line brightness distributions naturally lead to a non-uniform EW distribution.

The Fe\,{\sc xxv} EW map in Fig. \ref{fig:maps} shows an excess morphology similar to that of the surface brightness but extends further to larger radii in both northern and southern directions. Curiously, these extensions appear to originate from a particular region in the disc. These excesses show EW ranging from 0.2 keV to 0.6 keV and includes dX37 which shows an exceptionally large EW of $\sim 1.9$ keV. More interesting, however, are the small EWs observed in a number of segments. Some of them even have no Fe\,{\sc xxv} detection with the EW upper limits of 0.1 keV, despite of having source counts comparable to the other segments. If the diffuse emission is homogeneous gas, then individual segments are expected to have a constant Fe\,{\sc xxv} EW of 0.38 keV as measured in the whole diffuse emission spectrum (Table \ref{tab:feklines}). Assuming that the line counts follow a Poisson distribution, the observed EWs would be found in the 0.2-0.6 keV range at a probability of 90\%, given the observed counts (eight counts in the line, on average, was assumed). The fact that 17 segments have Fe\,{\sc xxv} EW $<0.2$ keV, as opposed to an expected 2 (equivalent to $\sim 5$\% of the total 38 segments), indicates that the observed EW variations are statistically significant. This spatially biased distribution of Fe\,{\sc xxv} means that a significant portion of the diffuse emission areas are void of Fe\,{\sc xxv} or have much weaker Fe\,{\sc xxv} emission than that seen in the average spectrum. This suggests a presence of a featureless continuum component of non-thermal emission. 

As for cold Fe line at 6.4 keV, its brightness CDFs are, in contrast, similar between north and south (Fig. \ref{fig:ecdf_surbri}). We note that there are two brightest segments, one each in the northern and southern segments, and the shape of the CDFs suggests that the total cold Fe flux is dominated by a small number of bright segments: the brightest five segments compose $\approx 40$\% of the total 6.4 keV line flux. The EW of the cold Fe line is similarly balanced between the north and south. There are 12 segments of $>90$\% detection of cold Fe line. Their EWs are clustered in the 0.16-0.3 keV range with an exception of 0.56 keV from dX28. If their distribution follows a normal distribution (the mean of 0.27 keV and standard deviation of 0.1 keV), two more segments with smaller EWs ($<0.16$ keV) are expected to join this distribution if the observation sensitivity was not limited. Therefore $\sim 40$\% (14/38) of the segments exhibit cold Fe lines with this moderately large EW of 0.1-0.5 keV. These segments spread over the whole ellipse region, suggesting a diffuse nature of the cold Fe line. We discuss its origin later. Because of the large difference in ionisation stage (and line excitation mechanisms) from Fe\,{\sc xxv}, it is no surprise to find different spatial distributions between their brightness and EWs.

\subsubsection{Remarkably strong Fe\,{\sc xxv} in dX37}

The spectrum of dX37 shows an Fe\,{\sc xxv} line with an exceptionally large EW of 1.9 keV (Fig. \ref{fig:map_values}). The Fe K band spectrum shows only the Fe\,{\sc xxv} line at $6.68^{+0.04}_{-0.01}$ keV but no 6.4 keV line (the 90\% upper limit of EW 0.35 keV). The 4-7.2 keV spectrum can be described by an {\tt apec} model with a temperature of $kT = 3.3\pm 0.9$ keV and metallicity of $1.6^{+1.0}_{-0.6} Z_{\sun}$, which is essentially that of Fe. The temperature is relatively low, compared to the other regions, resulting in the steep continuum slope ($\alpha\simeq 2.0$, Fig. \ref{fig:map_values}). Lowering a temperature leads to a larger Fe line EW at given metallicity (see Appendix A) but this pronounced line EW is well beyond the effect of varying temperature. The line EW is a factor of $\sim 5$ larger than that of the mean spectrum, and is comparable or even larger than those observed in the spectra of X-ray SNRs selected in \citet{iwasawa21}.  

As described in Sect. 3.3.1, this segment contains the transient source Xp11. However, as the image taken from the same time intervals used to construct the spectrum, that is the transient's quiescence state, shows no point-like source (Appendix C). No other discrete source, for example, a SNR, is present. No radio SNR is found in the segment either. The 4-6 keV continuum surface brightness is comparable to or even slightly lower than the adjacent segments, dX8 and dX9, located at the similar latitude from the disc. The 6.55-6.9 keV narrow-band image shows a flat distribution of counts within dX37, supporting the diffuse nature of the Fe\,{\sc xxv} emission.

An unlikely local metallicity enhancement could boost the line EW, but in the absence of any SN generating objects, such as a star cluster, that coincides with the region which lies $\sim 90$ pc above the molecular disc plane, it does not seem to be the case. Incidentally, the Fe metallicity found in this spectrum is close to $Z_{\rm Fe}\simeq 2~Z_{\sun}$ expected for the metal-rich hot gas with the best estimate of mass loading by \citet{strickland09}. This leads to a possibility that we might happen to see nearly pure emission from the hot gas component in this region with a minimal contribution of non-thermal emission which dominates everywhere else. This possibility is further explored below (Sect. 4).

\subsubsection{Relation between continuum slope and Fe\,{\sc xxv} strength}

\begin{figure}
\centerline{\includegraphics[width=0.34\textwidth,angle=0]{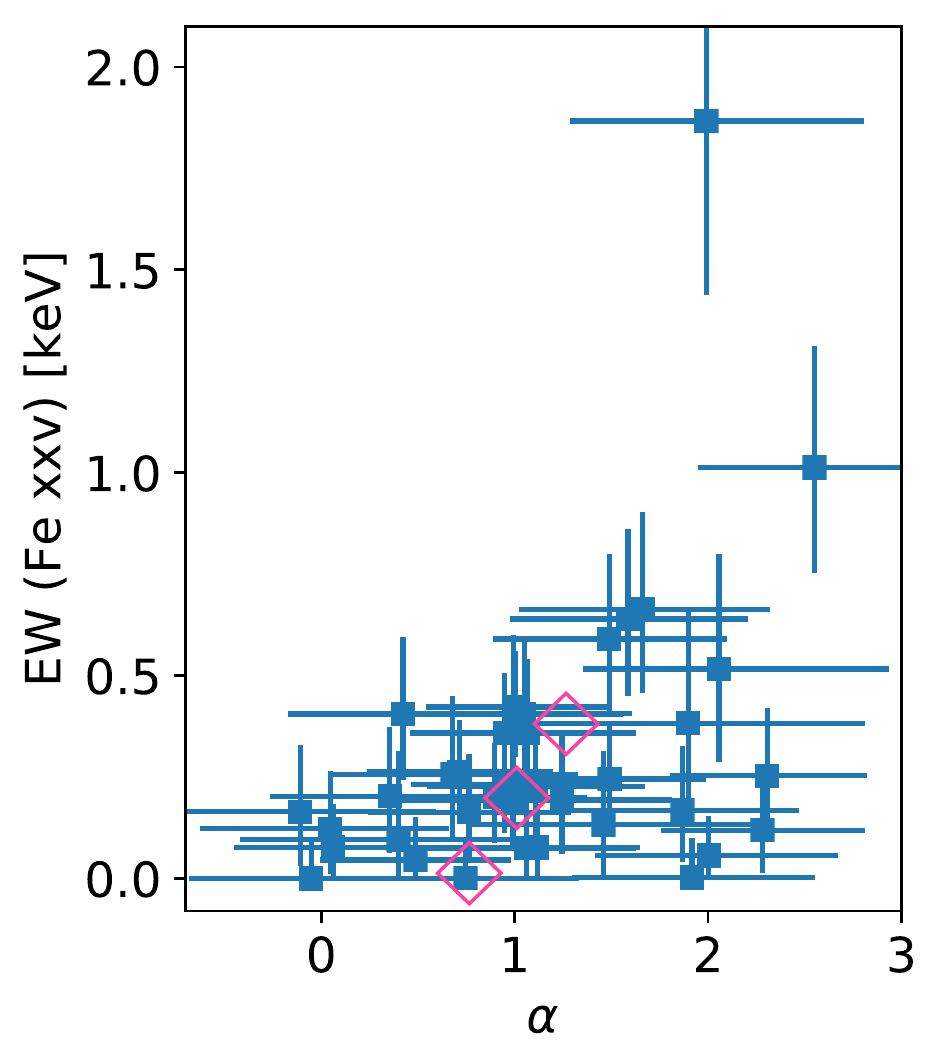}}
\caption{Plot of EW (Fe\,{\sc xxv}) against continuum slope $\alpha $ of the 38 region segments (in blue filled-squares). The three magenta diamonds indicate the points for the EW-binned spectra (Table \ref{tab:slopeEW}), where EW value of each point is the median of EWs measured in the spectra that went into each bin.}
\label{fig:slopeEW}
\end{figure}

Among the measured features, we noted a possible correlation between Fe\,{\sc xxv} EW and continuum slope (Fig \ref{fig:slopeEW}). Although the data are noisy and the correlation cannot be firmly established, six EWs of Fe\,{\sc xxv} larger than 0.5 keV, for example, occur only when the 4-6 keV slopes are $\alpha \geq 1.5$ while the median slope of all the spectra is $\tilde\alpha = 1.05$.

\begin{figure}
\centerline{\includegraphics[width=0.5\textwidth,angle=0]{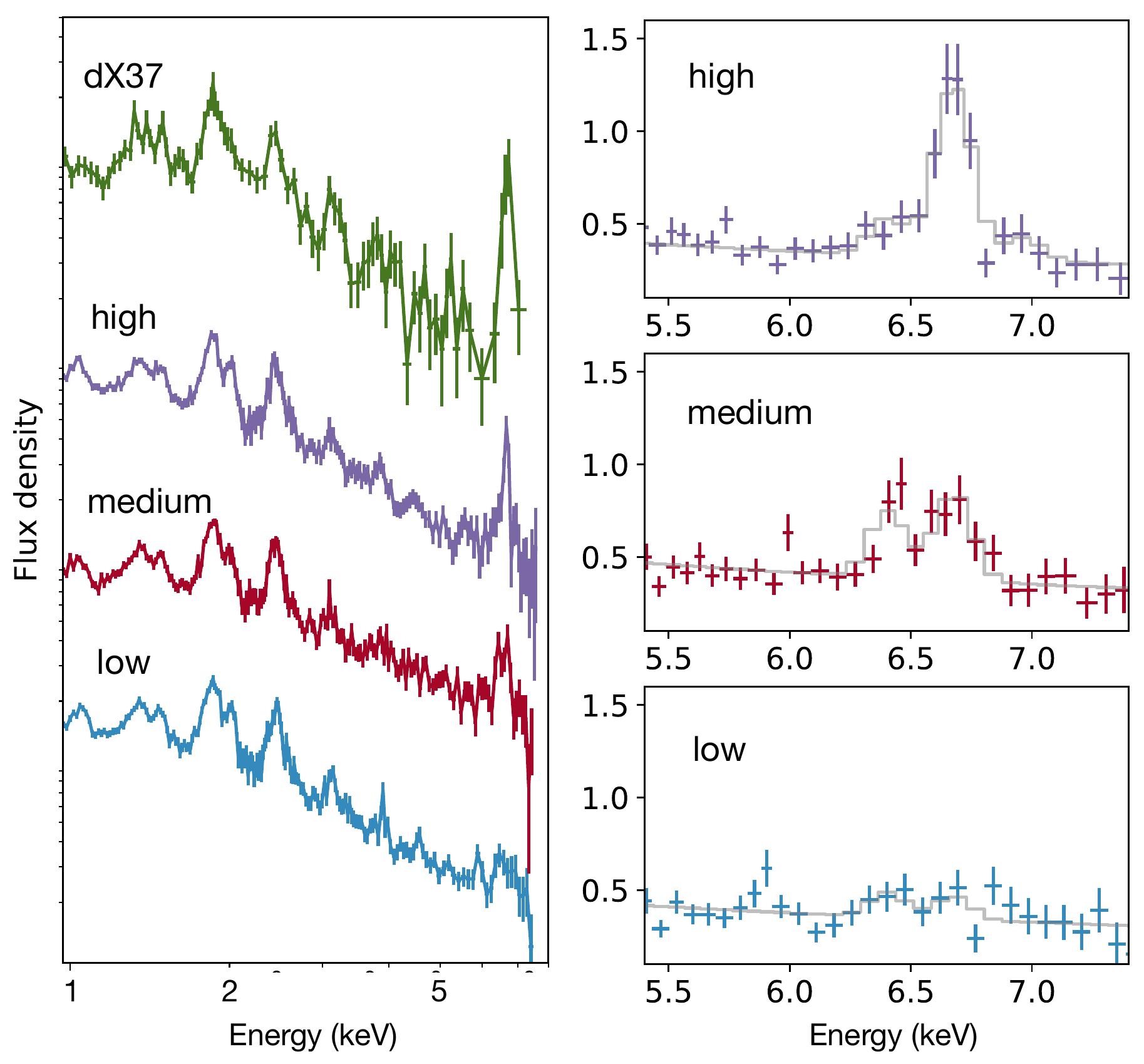}}
\caption{Spectral variations as a function of Fe\,{\sc xxv} EW. Left: The broad band spectrum of dX37, integrated spectra of Fe\,{\sc xxv} EW of EW $\geq 0.3$ keV, 0.15-0.3 keV, and $<0.15$ keV ('high', 'medium', and 'low' from the top to bottom). Each spectra has been multiplied arbitrarily for displaying purpose. Right: The Fe K band data of the high, medium, and low spectra in the left panel. The flux density is in units of $10^{-14}$ \ergpspsqcmpkeV. }
\label{fig:himidlo}
\end{figure}

Given the Fe\,{\sc xxv} EW variation (from $<0.1$ keV to 1.9 keV) examined above, an inclusion of non-thermal emission in addition to the thermal emission characterised by Fe\,{\sc xxv} can provide an explanation. In this hypothesis, varying composition of the two components across the diffuse emission region leads to the Fe\,{\sc xxv} EW variations. If the non-thermal emission has a harder spectrum than the thermal emission, a spectral slope of a composite spectrum varies as a function of Fe\,{\sc xxv} EW, as hinted above. We tested this by inspecting spectra binned into three Fe\,{\sc xxv} EW intervals, excluding dX37 with the exceptionally strong Fe\,{\sc xxv}. The three EW intervals were set as follows: 'low', EWs $\leq 0.15$ keV; 'medium', $0.15\,{\rm keV}< $ EWs $\leq 0.3$ keV; and 'high', $0.3 <$ EWs $\leq 1$ keV (Table \ref{tab:slopeEW}) and averaged spectra from the three EW bins were obtained. Their spectra, together with the dX37 spectrum as a comparison, are shown in Fig. \ref{fig:himidlo}. The spectra from the three bins have background-corrected counts of 0.9-7.5 keV: 59645 (4-7.5 keV: 2972), 56247 (3730) and 61650 (3113) for the low, medium and high EW bins. Their mean fluxes are comparable to one another as well as the quality of their spectra. The 5.4-7.4 keV band centred on the Fe K line complex of the spectra is also shown.

The three spectra are nearly identical in the soft X-ray band up to the Ca K line complex at $\sim 4$ keV, except for small variations at the lowest energies due to varying absorption across the region which correlates with the dust lanes visible in the optical image. This confirms homogeneous emission of the soft X-ray wind covers the entire hard-band diffuse emission region in foreground.

In contrast, the Fe K band spectra are dramatically different between them. The continuum slopes of the three spectra measured in the 4-6 keV band are given in Table \ref{tab:slopeEW}. The continuum steepens as increasing EW bins (overplotted also in Fig. \ref{fig:himidlo}). This trend continues with the even steeper slope found in the dX37 spectrum (Table \ref{tab:slopeEW}). This agrees with the idea that Fe\,{\sc xxv} EW is observed to be small when a contribution of non-thermal emission increases and vice versa. 

\begin{table}
  \caption{Fe\,{\sc xxv} EW sliced spectra.}
  \label{tab:slopeEW}
  \centering
  \begin{tabular}{cccrc}
  \hline\hline
Bin & EW(Fe {\sc xxv}) & $N$ & Counts & $\alpha $ \\
\hline
low & $\leq 0.15$ & 12 & 59645 & $0.84\pm 0.13$ \\
medium & 0.15-0.3 & 13 & 56247 & $1.02\pm 0.12$ \\
high & $>0.3$ & 12 & 61650 & $1.26\pm 0.14$ \\[5pt]
dX37 & 1.9 & 1 & 3865 & $2.0\pm 0.6$ \\
\hline
  \end{tabular}
  \tablefoot{EW of Fe {\sc xxv} are in units of keV. $N$ is the number of integrated segments. Counts are net source counts in the 0.9-7.5 keV band. The energy index $\alpha $ is the spectral slope measured in the 4-6 keV band. The same properties for the spectrum from dX37 are also given for comparison.}
  \end{table}

\subsection{Light distribution in 4-8 keV}

\begin{figure}
\centerline{\includegraphics[width=0.5\textwidth,angle=0]{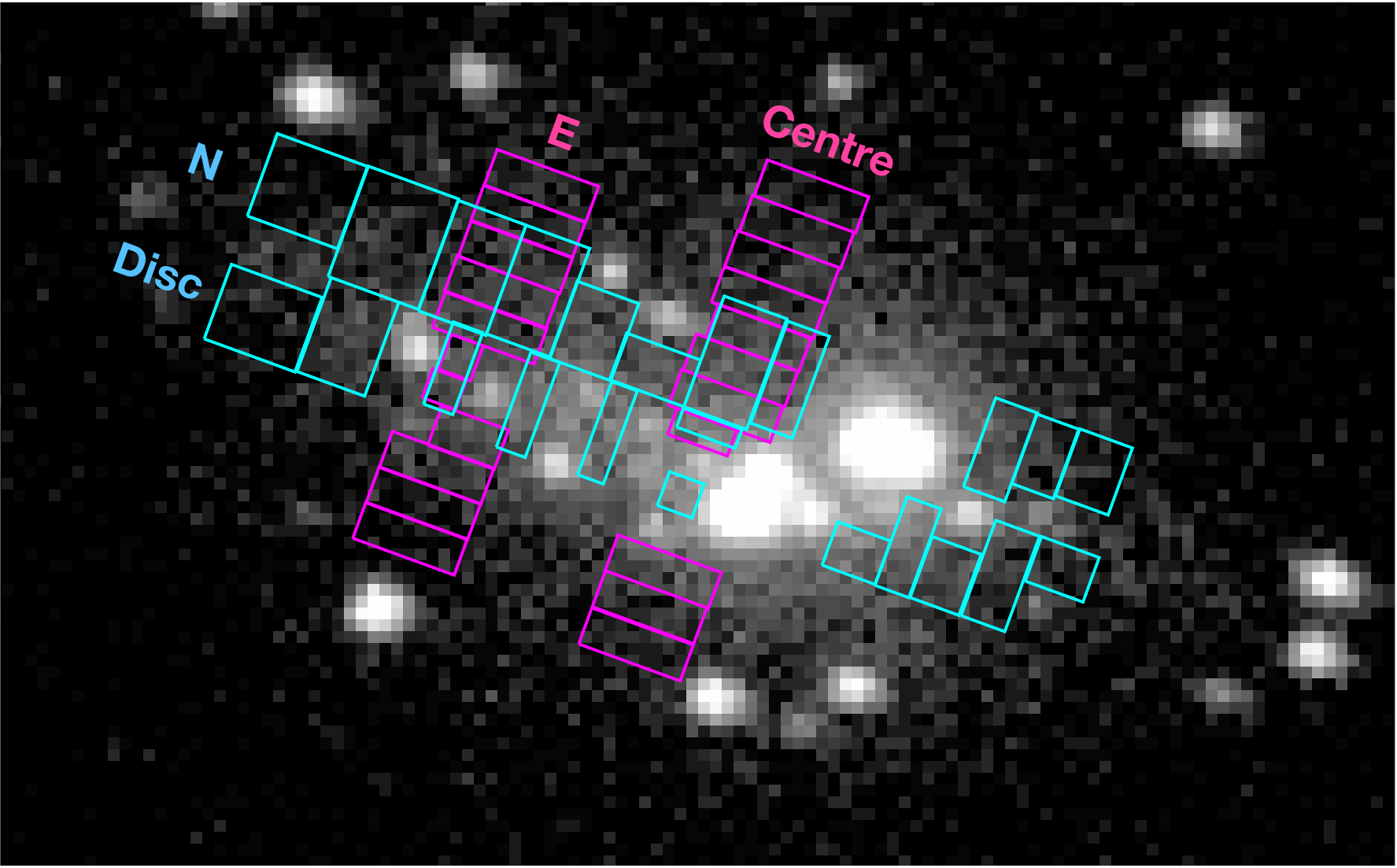}}
\caption{Image of the 4-8 keV emission overlaid by the two strips along the major axis: 'Disc' and 'N', and the two strips along the minor axis: 'Centre' and 'E'. }
\label{fig:sbprof_regions}
\end{figure}

To measure the approximate shape of the 4-8 keV diffuse light, we took projected light profiles along both the major and minor axes of the galactic disc as shown in Fig. \ref{fig:sbprof_regions}. Along the major axis (PA $= 70^{\circ}$), one strip labelled as 'major-Disc' goes through the molecular disc while the other strip, 'major-N', lies $\sim 70$ pc above the disc plane in the middle. The strips are inevitably disrupted by the discrete sources which were masked out. Two strips along the minor axis, 'minor-E' and 'minor-Centre', are also placed. The reference (zero) point of all the strips is the position of SN2008iz (RA = 09$^{\rm h}$55$^{\rm m}$51\fs55, Dec $= +69\degr 40\arcmin 45\farcs 8$, \citet{brunthaler09}), which is roughly coincides with the FIR emission peak (see Sect. 3.6). Background-corrected 4-8 keV counts were collected from each rectangular region and the surface brightness profiles along each strip was constructed.

\begin{figure}
\centerline{\includegraphics[width=0.5\textwidth,angle=0]{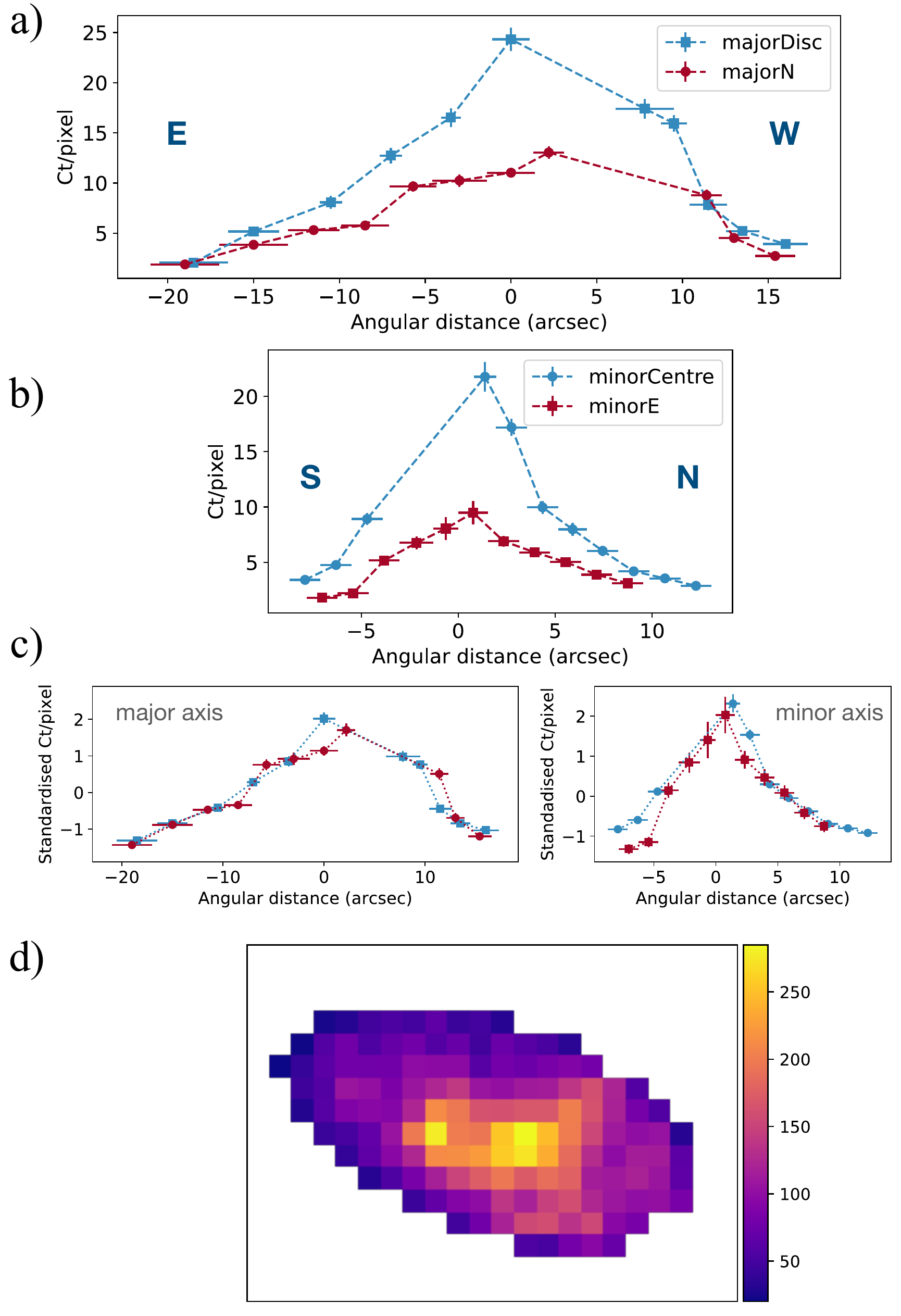}}
\caption{Light distribution in the 4-8 keV band. From top to bottom: a) The 4-8 keV projected light profiles along the major axis taken from the strips shown in Fig. \ref{fig:sbprof_regions}. b) The profiles along the minor axis. c) These two panels show the same profiles but standardised for a comparison of their shapes. d) The 4-8 keV diffuse emission image, constructed by two-dimensional interpolation from the point source masked image. Each pixel has an $\approx 2\times 2$ arcsec$^2$ size and the colour bar indicates the brightness scale in units of counts per pixel. }
\label{fig:sbprof}
\end{figure}

On inspecting those projected profiles, the brightness peak can be located at around the reference point. The light distribution along the major axis is relatively flat and skewed towards the east (Fig. \ref{fig:sbprof}a). Along the minor axis, the light distribution is nearly symmetric at the brighter inner part, as already seen in the 4-6 keV map (Fig. \ref{fig:maps}), but shows an enhanced extension towards north at the fainter brightness level (Fig. \ref{fig:sbprof}b). Along each axis, the profile shapes of the two strips are found to be similar, as shown in the standardised brightness plots (Fig. \ref{fig:sbprof}c), while their absolute fluxes differ.


Another purpose of creating these projected light profiles is to estimate a flux of the diffuse emission in the masked area of the clustered ULXs and other sources, Xp12 though Xp17. By interpolating the four profiles, the 4-8 keV flux in the area was estimated to be $\approx 3.3\times 10^{-13}$ \ergpspsqcm, by assuming the same counts-to-flux conversion factor obtained for the diffuse emission spectrum in Sect. 3.2. A two-dimensional interpolation can also be applied directly to the 4-8 keV image with the point source mask used for the 4-6 keV image in Fig. \ref{fig:e24e48img}. For estimating a missing light profile of the relatively large masked area, we adapted a wide grid-interval and thus the image was re-binned by four pixels before applying the interpolation ({\tt interpolation.griddata} from Scipy) with the 'cubic' method. The resulted interpolated image is shown in Fig. \ref{fig:sbprof}d. The missing flux of the masked area, computed from this map, agrees with the above estimate.

\subsection{Diffuse emission luminosity}

In addition to the ULX area, the expected diffuse flux from the other masked areas of discrete sources was computed by applying the same conversion factor and it is found to be $0.9\times 10^{-13}$ \ergpspsqcm. Adding these estimated fluxes from the masked areas to the $7.2\times 10^{-13}$ \ergpspsqcm\ of the diffuse emission spectrum (Sect. 3.2), the total 4-8 keV flux of diffuse emission in the ellipse in Fig. \ref{fig:e24e48img} is thus $1.1\times 10^{-12}$ \ergpspsqcm, which corresponds to the 4-8 keV luminosity of $1.7\times 10^{39}$ \ergps. 

This luminosity, however, contains some contribution of unresolved discrete sources, that is, X-ray binaries and SNRs. We estimated their contributions using the cumulative luminosity distribution, $N(>L)$, which can be approximated by a power-law form ($\propto L^{-\gamma }$). For X-ray binaries in star-forming galaxies, $\gamma\simeq 0.6$ has been found \citep{colbert04,grimm03,mineo12}. As $\gamma < 1$ implies that brighter sources dominate, fainter, unresolved X-ray binaries have a minor contribution to the observed diffuse emission luminosity anyway. The $N(<L)$ of the X-ray binaries detected by \citet{iwasawa21} down to $1\times 10^{-14}$ \ergpspsqcm\ in the 4-8 keV band, below which the detection completeness declines rapidly, within the diffuse emission region (excluding the far more luminous ULXs) also show $\gamma\sim 0.6$. Using this slope, we estimated the 4-8 keV luminosity of unresolved X-ray binaries to be $2.0\times 10^{38}$ \ergps. The $N(<L)$ slope for X-ray SNRs appears to have a similar value $\gamma = $0.6-0.7, based on the seven X-ray SNRs \citep{iwasawa21}. The 4-8 keV luminosity from unresolved X-ray SNRs was then found to be $0.4\times 10^{38}$ \ergps.

A further correction needed is for photons spilled over from the masked discrete sources. About 10\% of each discrete source counts in the PSF wing fall outside the masked aperture, on average. This is estimated to be $\sim 0.6\times 10^{38}$ \ergps. Taking all these into account, the net diffuse emission luminosity in the 4-8 keV band is found to be $1.4\times 10^{39}$ \ergps, in which $1.3\times 10^{39}$ \ergps\ comes from the continuum (approximately 9\% of the 4-8 keV luminosity originates in Fe\,{\sc xxv}). Unless this luminosity solely originates in thermal emission described by the {\tt apec} fit and in a spheroid as marked by the ellipse in Fig. \ref{fig:e24e48img}, the mean electron density would be $n_e\sim 0.6$ cm$^{-3}$, as the volume of the spheroid is $\sim 7.9\times 10^7$ pc$^{3}$.

\citet{strickland07} estimated the 2-10 keV continuum luminosity of $4.4\times 10^{39}$ \ergps, which corresponds to $2.0\times 10^{39}$ \ergps\ in the 4-8 keV band. Our estimate is lower than theirs by about one third. The difference might be attributed to more rigorous removal of discrete sources, including the X-ray SNRs, with the much deeper imaging.

\subsection{X-ray, FIR, and radio images}


\begin{figure*}
\centerline{\includegraphics[width=0.8\textwidth,angle=0]{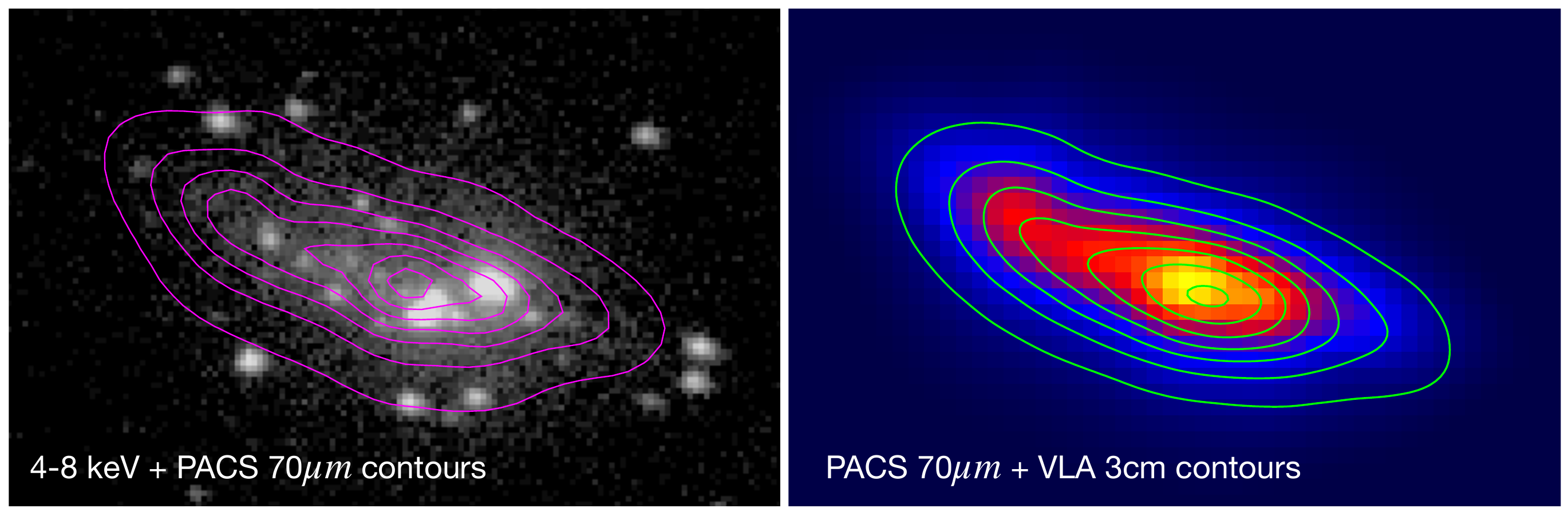}}
\caption{Diffuse X-ray, FIR, and radio image comparison. Left: The 4-8 keV {\it Chandra} image in grey-scale overlaid by the {\it Herschel}-PACS 70 $\mu $m FIR contours in magenta. Right: The PACS 70 $\mu $m image overlaid by {\it VLA} 3-cm continuum \citep{adebahr13} contours in green. The beam sizes of the PACS and the {\it VLA} images are $\sim 5.5$ arcsec and $7.6\times 7.3$ arcsec, respectively. Both contours are drawn in seven levels divided in linear scale with the lowest contour corresponding to 10\% of the peak brightness in the respective images.}
\label{fig:radioFIR}
\end{figure*}

The spatial variation of Fe\,{\sc xxv} investigated in Sect. 3.3 led us to consider the hypothesis that the 4-8 keV diffuse emission might be a composite of thermal and non-thermal emission. The non-thermal emission, in this case, arises from inverse Compton scattering of FIR photons by (sub-GeV) cosmic ray electrons \citep{hargrave74}. Ample FIR emission is radiated from the dusty starburst region while the cosmic rays are accelerated by SNRs that are abundant in the same region. The electron population of the latter is represented by the synchrotron radio emission. Therefore if this inverse Compton emission is responsible for the diffuse X-ray emission, its morphology is expected to match that at the FIR and radio wavelengths.

We used the FIR image (Fig. \ref{fig:radioFIR}) taken by the {\it Herschel} PACS \citep[][data were obtained through NED]{bendo12} for a comparison. It provides a $70\thinspace\mu $m image at the angular resolution of $\sim 5\farcs 5$ (PACS Handbook 2019, version 4.01 ESA). A similar image at $59\,\mu $m (the beam FWHM of $5\farcs 5$) has been also taken by {\it SOFIA} \citep{jones19}. We used a radio continuum image (Fig. \ref{fig:radioFIR}) taken by the {\it VLA} at 3~cm \citep{adebahr13} with D-configuration with a matching beam size ($7\farcs 6\times 7\farcs 3$) to the FIR image. The radio continuum at 3-cm represents the high-frequency part of the synchrotron emission with a power-law spectrum $\nu^{-0.7}$ in the starburst region while the electron population responsible for the Compton scattering are attributed to those at lower frequencies \citep[e.g.][]{strickland07}. The power-law spectrum shows a turnover below 1 GHz due to free-free absorption \citep{adebahr13}, meaning that any image taken at frequencies lower than 1 GHz is affected by the absorption which suppresses the brightness in the high obscuration region. For this reason, the high-frequency image is expected to represent the intrinsic distribution of the cosmic ray electron energy density.

The FIR and radio images are, unsurprisingly, very similar (Fig. \ref{fig:radioFIR} Right). Their image morphology agrees well with that of the 4-8 keV diffuse emission (Fig. \ref{fig:radioFIR} Left; see also Fig. \ref{fig:sbprof}d). The matching morphology between the three band images is encouraging for the inverse Compton scattering origin for non-thermal emission. A likely contribution of this non-thermal component to the 4-8 keV continuum luminosity is further discussed in Sect. 4.   

\subsection{Galactic centre source}

\begin{figure}
\centerline{\includegraphics[width=0.42\textwidth,angle=0]{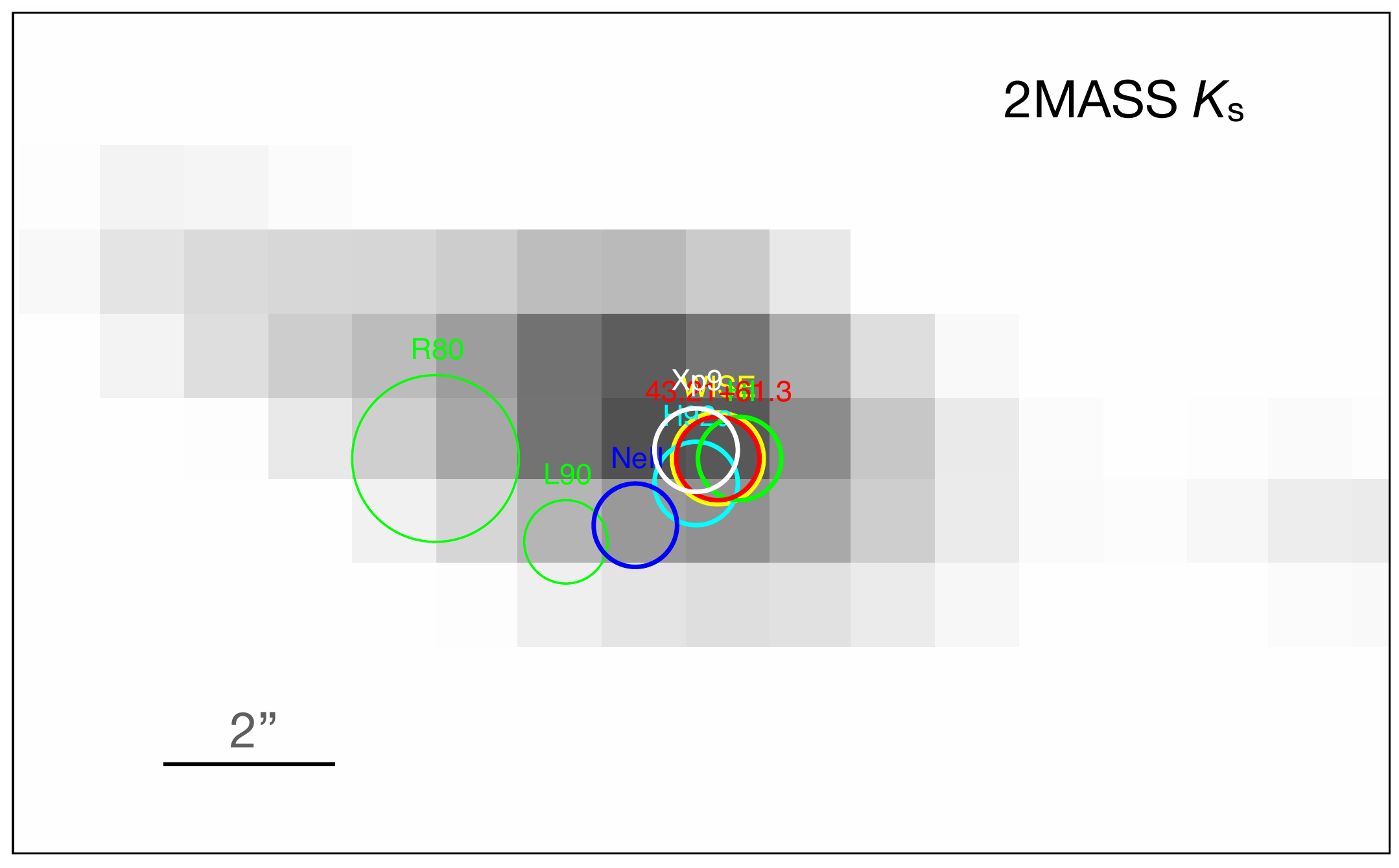}}
\caption{$K_{\rm s}$ band image of the M82 nuclear peak emission obtained from 2MASS in grey scale (fainter emission is cut out), overplotted by positions of the X-ray source, Xp9 (white), radio source 43.21+61.3 \citep[red,][]{rodriguez-rico04}, H\,{\sc i} kinematic centre \citep[green][]{weliachew84}, Ne\,{\sc ii} kinematic centre \citep[blue,][]{achtermann95}, H92$\alpha $ ring centre \citep[cyan,][]{rodriguez-rico04}, WISE peak \citep[yellow,][]{jarrett19}, and two previous 2.2\,$\mu $m peak measurements \citep[R80,L90:][]{rieke80,lester90}. The positional error is $\sim 0.5\arcsec$ for each measurement, which is represented by the circle, apart from the 2.2\,$\mu $m position of \citet{rieke80}. The image orientation is north up, east to the left. }
\label{fig:galcentres}
\end{figure}

The 2.2\,$\mu $m peak \citep{lester90} has often been assumed to be the galactic centre position of M82,
as the near-IR emission is a good tracer of stellar population and relatively robust against obscuration. However, the $K$-band emission has a complex structure \citep[see e.g.][]{alonso-herrero03} and the large X-ray absorbing column of \nH $\sim 10^{23}$ \psqcm\ suggests that obscuration could affect the $K$-band peak position. Instead, there are a few measurements of a kinematic centre determined by H\,{\sc i} \citep{weliachew84}, [Ne\,{\sc ii}] \citep{achtermann95}, and radio H92$\alpha $ line \citep{rodriguez-rico04} observations.
We noted that the X-ray source, Xp9, detected in \citet{iwasawa21}\footnote{The source position is 09h55m52.0s, +69d40m47.3s, There was a typo in the Dec. in \citet{iwasawa21}. This is the correct position of Xp9.} is located at the radio kinematic centre, and has a relatively faint radio counterpart. The radio source is catalogued as '43.21+61.3' in \citet{rodriguez-rico04} who observed with {\it VLA} at 8.3 GHz and 43 GHz and detected with 4 mJy in both frequencies. It is tentatively classified as a H~{\sc ii} region because of the flat spectrum. These positions, including the previously measured 2.2\,$\mu$m peak positions \citep{rieke80,lester90} and the {\it WISE} position \citep[][WXSC]{jarrett19}, are plotted over the 2MASS $K_{\rm s}$ band image (Fig. \ref{fig:galcentres}, retrieved from the 2MASS Image Service of IRSA). The positions of Xp9 and the radio source agree with the kinematic centre measurements and the {\it WISE} position, and is $\la 1\arcsec $ off to the west from the 2MASS peak. 
This positional coincidence with the kinematic centre, albeit being tentative, offers a possibility that Xp9 might be a low-luminosity AGN sitting at the galactic centre of M82. This possibility is examined in Sect. 4.7.

\section{Discussion}

We discuss the origin of the diffuse 4-8 keV emission which appears to be composed of multiple components of distinct origins.
Three lines of evidence below suggest the presence of significant non-thermal emission in the 4-8 keV diffuse emission: 1) Substantial portions of the diffuse emission region exhibit weak or no Fe\,{\sc xxv} in the spectrum (Sect. 3.3.2); 2) The continuum hardening with decreasing Fe\,{\sc xxv} EW (Sect. 3.3.4); and 3) The morphological resemblance between the X-ray, FIR and radio images (Sect 3.6).
Based on this argument, we present a spectral decomposition of the diffuse X-ray emission that includes inverse Compton emission, followed by a discussion of the IC emission, metal-rich hot gas, its feedback on ISM and relation to the X-ray SNRs, the 6.4 keV Fe line, and the galactic centre source.

\subsection{Composition of the 4-8 keV diffuse emission}

Our hypothesis is that the diffuse emission is inverse Compton emission and soft X-ray wind emission, added to the hot gas of $kT\sim 5$ keV which occupies specific regions where the Fe\,{\sc xxv} line is detected. Based on this picture, we tried to assess the components of the 4-8 keV diffuse emission. We have to consider at least three continuum components: 1) IC: the inverse Compton emission; 2) Th1: the high-energy tail of the soft X-ray wind emission; and 3) Th2: the thermal emission giving rise to the Fe\,{\sc xxv} line. As it is not practical to fit all the parameters of such a complex model, we assumed reasonable spectral parameters for each component as described below.

The inverse Compton emission component (IC) is modelled by a power-law. As the spectral slope is expected to be approximately the same as the radio synchrotron spectrum, we assumed the energy index of $\alpha = 0.7$. This component has the hardest spectrum among the three components.

The soft X-ray wind emission (Th1) is modelled by {\tt apec} with $kT=1.2$ keV. This component, which dominates in the soft X-ray band, should still have a significant contribution in continuum emission at energies around 4-5 keV. The assumed temperature (and also the adopted \nH\ value) was that obtained for the S K band (Fig. \ref{fig:spfit}). We believe this choice is adequate for the soft X-ray wind emission spectrum at energies of Ar K and Ca K emission and above 4 keV for the reason described in Sect. 3.2.1. The metallicity is assumed to be $1~Z_{\sun}$ for the following reasons. It has to be larger than $0.6~Z_{\odot}$ inferred from the soft X-ray lines (Fig. \ref{fig:spfit}) to compensate the extra continuum contribution of the IC component. With this temperature, the spectrum has high-ionisation Fe K emission including Fe\,{\sc xxv}, centred at 6.63 keV (see Appendix A) and it should not overproduce the Fe\,{\sc xxv} feature of the 'low' spectrum, in which the hot thermal emission (as described below) contribution is supposed to be negligible.

The hot thermal emission (Th2) is modelled by {\tt apec} with a variable temperature around $kT=5$ keV. The Fe metallicity is a key quantity: since this component is the primary source of the observed Fe\,{\sc xxv} line (and also Fe\,{\sc xxvi}). If the Fe metallicity were known, the continuum strength would automatically be determined. However, we do not know it a priori. Since the reverse is also true and the continuum level, which is driven by the IC component strength, mediates the Fe metallicity, we exploited this route to constrain the Fe metallicity, as below. \citet{strickland07} derived the original SN ejecta to have Fe metallicity of $5~Z_{\sun}$ through a Starburst99 \citep{leitherer99} simulation. This value can be considered as the upper bound and coincides also with the upper bound of the {\tt apec} model parameter. The lower bound is $1~Z_{\sun}$, since the Th1 component is mass-loaded further to the hot fluid by entraining ISM and therefore has lower metallicity than that of Th2.

All these components are assumed to be modified by cold absorption of \nH $= 2\times 10^{22}$ \psqcm. The two additional lines at 4.5 keV and 6.4 keV, which are required to describe the data (Sect. 3.2) but not of our interest here, were added as narrow Gaussians. 

The three Fe\,{\sc xxv} EW sliced spectra in Sect. 3.3.4 (Fig. \ref{fig:slopeEW} and Table \ref{tab:slopeEW}) have continuum slopes that depends on their line EWs. The dX37 spectrum with the exceptionally large Fe\,{\sc xxv} EW follows the same trend. We assessed the composition of the three continuum components that can reproduce their varying spectral slopes, which, in turn, constrains the Fe metallicity of Th2. The 'low' Fe\,{\sc xxv} EW spectrum can be considered to have a negligible contribution of Th2, as its weak Fe\,{\sc xxv} line can be accounted for solely by Th1. On the contrary, the dX37 spectrum should be dominated by Th2. We fitted the three EW-sliced spectra and dX37 spectrum in the 4-7.2 keV band jointly, leaving the normalisation of each component and the temperature and (Fe) metallicity of Th2 as free parameters.

\begin{figure}
\centerline{\includegraphics[width=0.35\textwidth,angle=0]{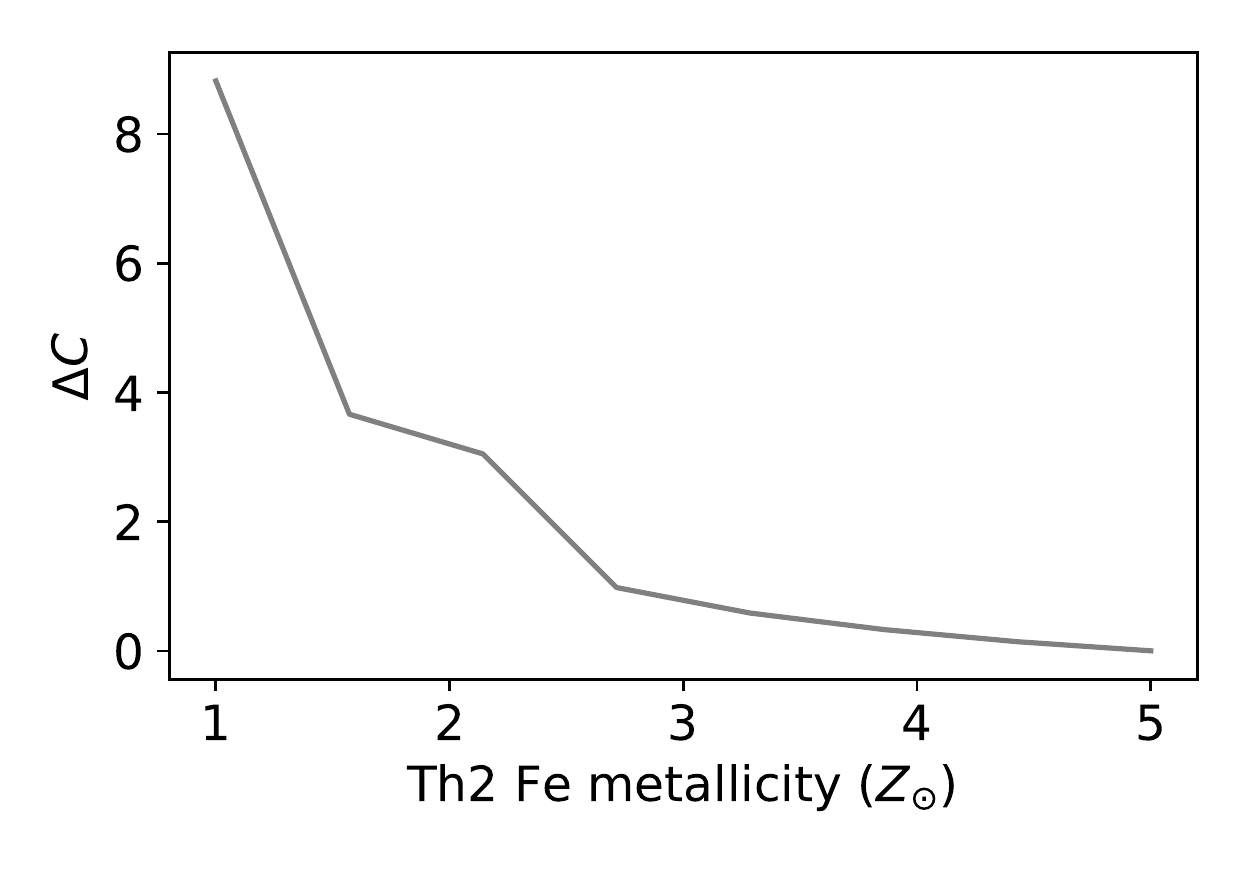}} 
\caption{C-statistic curve for the Fe metallicity of Th2 in the joint fit of the three EW-sliced spectra and dX37 spectrum with the three  continuum component model.}
\label{fig:zfe}
\end{figure}

This fit gives a good description of the data with C-stat = 785.3 with 774 bins (754 dof). The temperature was found to be $kT = 5.0\pm 0.8$ keV. The best-fit Fe metallicity was driven to the highest bound of $5~Z_{\sun}$ but the likelihood remains similar above $2.5~Z_{\sun}$ (Fig. \ref{fig:zfe}). This can be compared with the Fe metallicity of $\sim 2~Z_{\sun}$, which would be expected from the best-estimate of the mass-loading factor of 2.7 by \citet{strickland09}.

\begin{figure}
\centerline{\includegraphics[width=0.35\textwidth,angle=0]{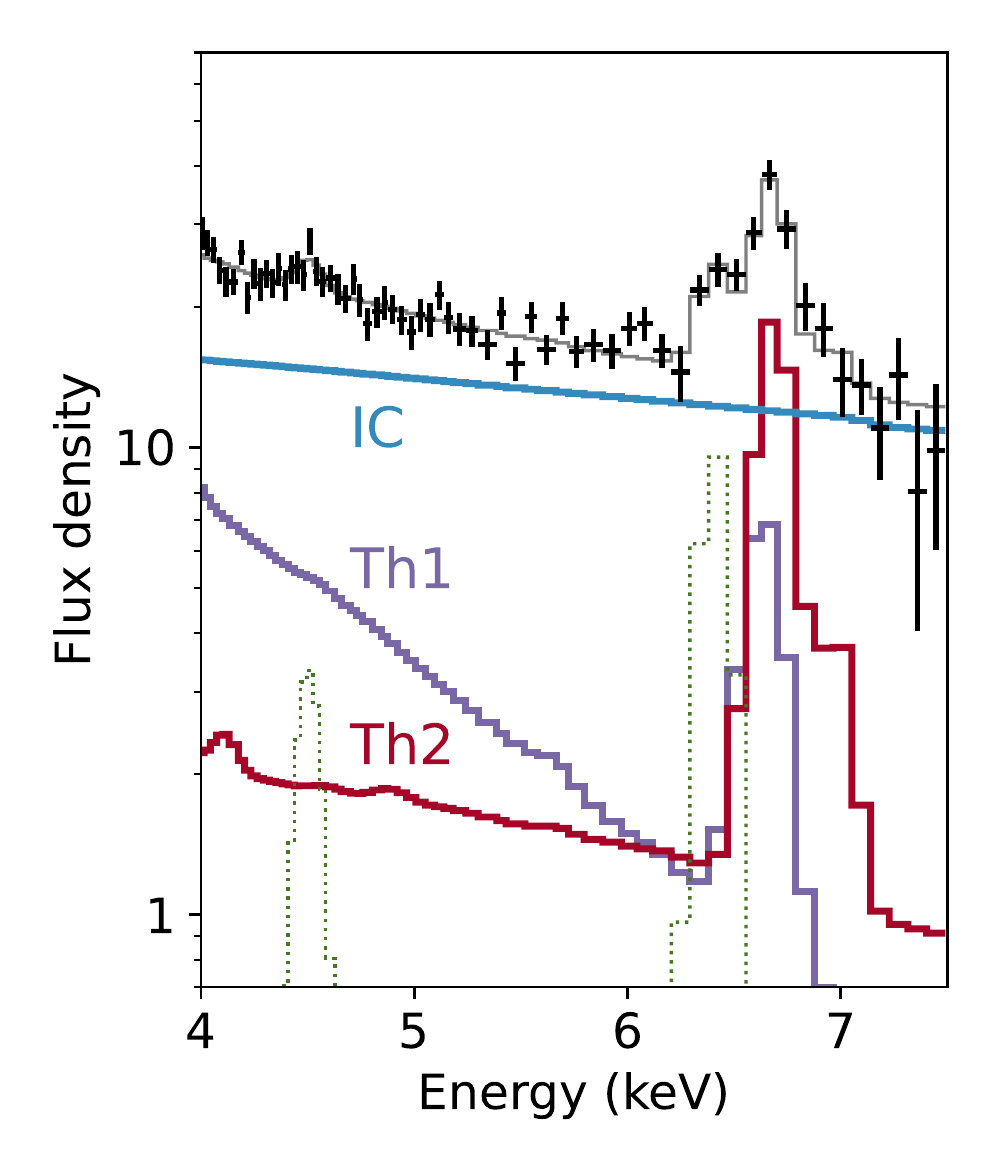}}
\caption{Same spectrum shown in Fig. \ref{fig:fnbb} with the best estimate of the three continuum components, IC (blue), Th1 (purple) and Th2 (red) assuming the Fe metallicity of Th2 to be $3.8~Z_{\sun}$. Two additional lines at 4.5 keV and 6.4 keV are also plotted in dotted lines. The flux density is in units of $10^{-14}$ erg\,s$^{-1}$\,cm$^{-2}$\,keV$^{-1}$.}
\label{fig:compositions}
\end{figure}

We computed 4-8 keV fluxes of the IC, Th1 and Th2 components in the four EW-sliced spectra, adopting the Th2 Fe metallicity of $3.8~Z_{\sun}$, which is the probability-weighted mean over the expected metallicity range of 1-5 $Z_{\sun}$. Their proportions of the total 4-8 keV flux in each spectrum are given in Table \ref{tab:compositions} (where the contribution of the 4.5 keV and 6.4 keV lines are left out). With the same set of spectral components, a spectral decomposition for the total diffuse emission spectrum (Fig. \ref{fig:fnbb}) is illustrated in Fig. \ref{fig:compositions}. The relative contributions of the respective components are also given in Table \ref{tab:compositions}. 

We note that the Fe K line at 6.67 keV is actually a sum of the lines from Th1 and Th2. The Fe K feature of Th1 is centred at 6.63 keV, consisting of Fe\,{\sc xxv} and weaker lines originating in lower ionisation stages, given the temperature of $kT=1.2$ keV. This Th1 contribution probably shifts the line centroid and broadens the width to the observed 6.67 keV and $\sigma = 0.05$ keV (Table \ref{tab:feklines}) from 6.68 keV and $\sigma = 0.03$ keV, expected from a $kT=5$ keV plasma.

The IC flux correlates weakly with the Fe metallicity via the Th2 continuum flux. In the probable 2.5-5 $Z_{\sun}$ range, the IC component carries 66-70\% ($\pm 3$\%) of the 4-8 keV diffuse emission luminosity, which corresponds to (0.9-1) $\times 10^{39}$ \ergps. This, however, contains the contribution of unresolved X-ray binaries with a similar spectral shape, which was estimated to be $0.2\times 10^{39}$ \ergps\ (Sect. 3.5). Correcting for this, the net 4-8 keV luminosity of inverse Compton emission is (0.7-0.8) $\times 10^{39}$ \ergps.

\begin{table}
  \caption{4-8 keV flux compositions.}
  \label{tab:compositions}
  \centering
  \begin{tabular}{lccc}
  \hline\hline
Bin & IC & Th1 & Th2 \\
\hline
low & $0.82\pm 0.04$ & $0.16\pm 0.02$ & 0.003 ($<0.025$) \\
medium & $0.74\pm 0.04$ & $0.13\pm 0.02$ & $0.09\pm 0.02$ \\
high & $0.55\pm 0.04$ & $0.17\pm 0.02$ & $0.25\pm 0.03$ \\
dX37 & 0.16 ($<0.33$) & $0.17\pm 0.09$ & $0.67\pm 0.17$ \\[5pt]
Total & $0.69\pm 0.03$  & $0.14\pm 0.02$ & $0.14\pm 0.02$ \\
\hline
  \end{tabular}
  \tablefoot{Proportions of the IC, Th1 and Th2 components in the 4-8 keV flux when the EW-sliced spectra, dX37 and total diffuse emission spectrum are fitted by the three-component model with the Th2 Fe metallicity of $3.8~Z_{\sun}$.}
  \end{table}

\subsection{Inverse Compton emission}

The parameters for estimating the energy loss of cosmic ray electrons due to inverse Compton scattering were obtained by \citet{adebahr13}, using their radio data, combined with the FIR data from {\it AKARI}. 
They agree with those used by \citet{persic08} who predicted a 2-10 keV flux of IC X-ray emission, which assumed that the entire FIR emission (measured with {\it IRAS}) comes from radii $\leq 300$ pc. However, since some fraction extends out, as shown in the {\it AKARI} \citep{kaneda10} and {\it Herschel} \citep{contursi13} images, a correction is needed to obtain the FIR energy density for the X-ray diffuse emission region. The $L_{\rm FIR}$ of the central $<1\arcmin $ in radius is $2.4\times 10^{43}$ \ergps, estimated using the \citet{helou88} formula with $S_{60}=630$\,Jy and $S_{100}=529$\,Jy, measured with {\it AKARI}-FIS \citep{kaneda10}. We found, from the PACS $70\,\mu$m image, that $\approx 60$\% of this luminosity comes from the ellipse of the X-ray diffuse emission. Weighting with the FIR emission distribution, the mean FIR energy density within the ellipse was estimated to be $U_{\rm FIR}\sim 0.82\times 10^{-9}$ erg\,cm$^{-3}$.
\citet{adebahr13} derived the magnetic field strength of $B=98\,\mu $G of the M82 central region, using the revised equipartition formula of \citet{beck05}. It is slightly smaller than $106\,\mu $G of \citet{persic08} but has an uncertainty of $\pm 30$\% due to the assumed proton to electron number density ratio. The slope of the radio spectrum is assumed to be $\alpha = -0.7$, as in Sect. 3.6.
This corresponds to the cosmic ray electron spectrum in terms of Lorentz factor $\gamma $: $N_{\rm e}(\gamma)\propto\gamma^{-2.4}$.
FIR photons at $70\,\mu $m, for example, are up-scattered to the 4-8 keV range by $\gamma\sim 500$-700. Following \citet{persic08} with these conditions, the expected 4-8 keV IC luminosity is found to be $0.6\times 10^{39}$ \ergps. This agrees with the decomposed IC emission luminosity found above.

\subsection{Metal-rich hot gas}

Even given its small contribution to the continuum emission, the thermal emission from the hot gas of $kT\simeq 5$ keV is the source of the majority of Fe\,{\sc xxv} line, as shown above, and can be traced by the enhanced Fe\,{\sc xxv} EW (Fig. \ref{fig:maps}). The map shows that the presence of the hot gas component is limited within a small distance from the starburst disc and seen to be much stronger on the northern side, suggesting an asymmetric ejection of hot gas.

A northern asymmetry in outflow structures have also been observed in other wavelengths. In the mid-infrared (mid-IR) image taken by the {\it Subaru}-COMICS with the [Ne\thinspace {\sc ii}] filter at $12.8\thinspace\mu $m, a few wisps are seen only on the northern side, extending from the starburst disc \citep{gandhi11}.
The same features are noted also in radio. Fig. \ref{fig:jvlaNeII} shows the deep {\it JVLA} image of the central part of M82 at 3-GHz, taken on 2015 August 16. The overlaid COMICS contours shows the coincidence of the northern wisps between mid-IR and radio. \citet{wills99} noted these radio counterparts in the earlier {\it VLA} images and identified four galactic chimneys, based on the emission void to the south of the radio SNR 41.95+57.5.
Here, we refer to a chimney as a collimated structure of a super bubble breaking out of the galactic disc. The primary energy source here is a sequence of SNe produced by a starburst and its formation has been formulated by earlier study \citep{tomisaka86,maclow88,norman89}.
The mid-IR and radio emission voids, embraced by the wisps, correspond to chimneys that are expected to be filled by hot gas \citep{tomisaka88}.

\begin{figure*}
\centerline{\includegraphics[width=0.65\textwidth,angle=0]{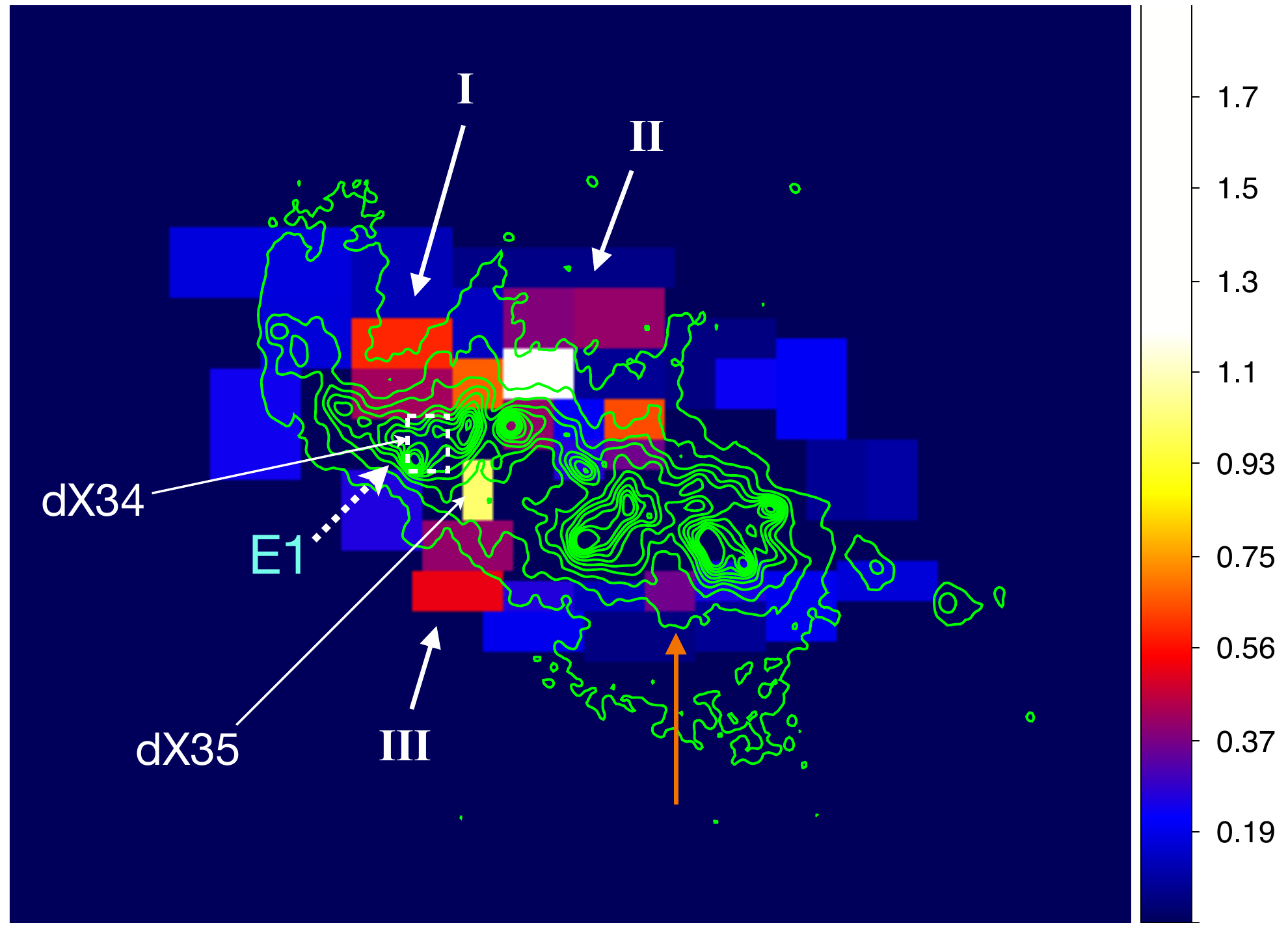}}
\caption{Fe\,{\sc xxv} EW map overlaid by the {\it Subaru}/COMICS mid-IR [Ne\,{\sc ii}] contours. The colour bar shows the Fe\,{\sc xxv} EW in keV. The white arrows indicates the three Fe\,{\sc xxv}-enhanced regions I, II, and III, which appear to fill the chimneys and to be connected to the mid-IR knot 'E1' \citep{achtermann95}. The orange arrow indicates the chimney with an expanding bubble identified around the SNR 41.95+57.5 by \citet{wills99}. The two region-segments dX34 and dX35, discussed in the text, are also indicated.}
\label{fig:hotEW+neii}
\end{figure*}

Fig. \ref{fig:hotEW+neii} shows that the Fe\,{\sc xxv} EW enhancement regions, as indicated by arrows I, II, and III. They appear to fill the mid-IR voids and to be traced back to the mid-IR knot 'E1' of \citet{achtermann95}.
This can be understood that the metal-rich hot gas produced in the E1 knot was channelled into the chimneys and fills their interior.

The mean density of the hot gas is found to be $n=0.09$ cm$^{-3}$. We note that if the hot gas does not fill the full diffuse emission volume, the density increases by $f^{-1/2}$, where $f$ is the filling factor.
The pressure of hot gas with $kT=5$ keV is then log $P\approx 6.7$ [$K$~cm$^{-3}$]. The hot gas of this temperature has a thermal velocity ($\sim 3000$ km~s$^{-1}$) much larger than the galaxy's escape velocity ($\sim 460$ km~s$^{-1}$, \citet{strickland09}) and, consequently, flows out. However, as the dense molecular gas in the starburst disc has a comparable pressure, log $P\sim 6.5$ [$K$~cm$^{-3}$], measured by \citet{naylor10}, the gas would flow out only in a favourable direction set out by the chimneys.

\subsection{Effects of supernova feedback}
\begin{figure}
\centerline{\includegraphics[width=0.5\textwidth,angle=0]{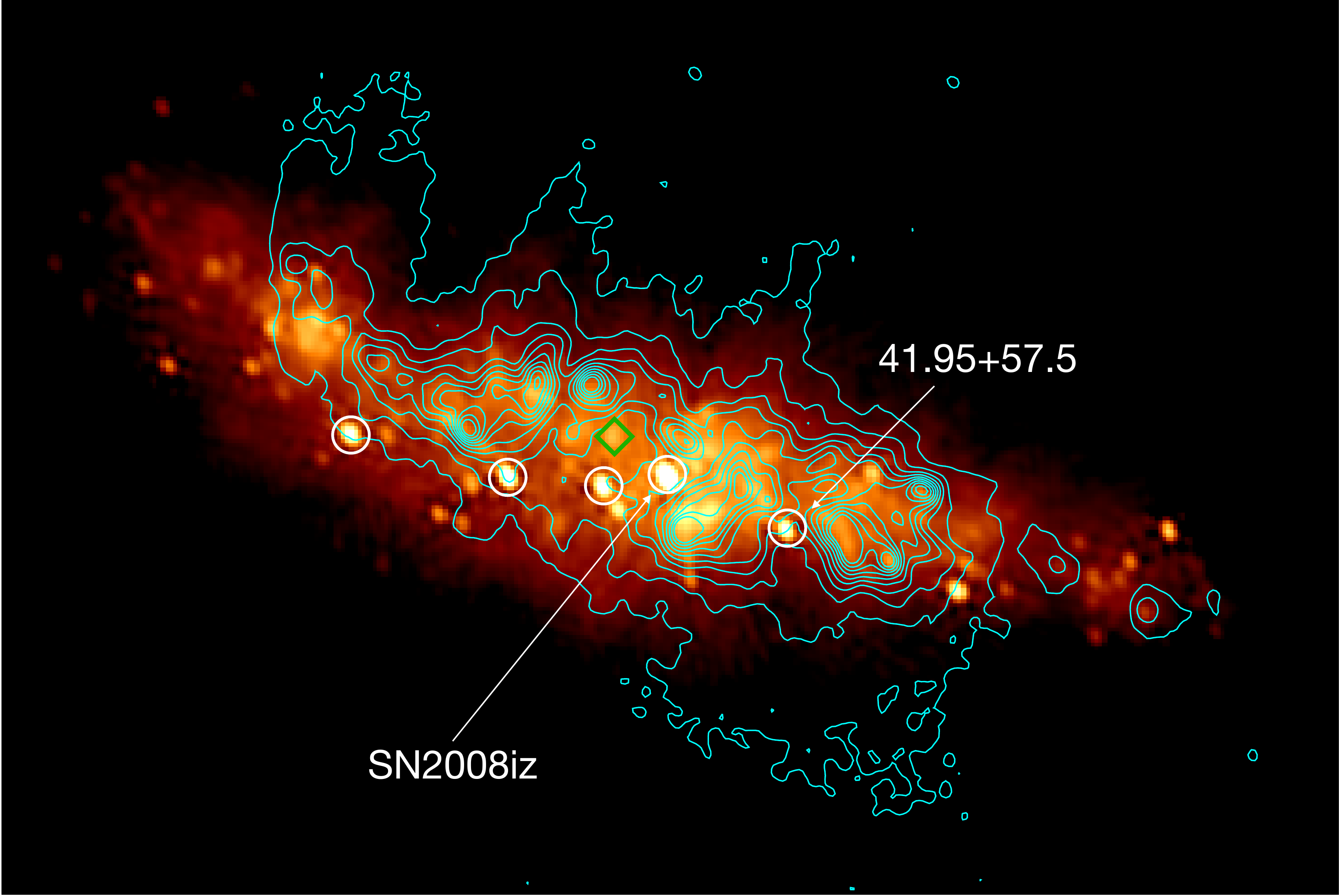}}
\caption{{\it JVLA} map at 3 GHz overlaid by the {\it Subaru}-COMICS [NII] contours. Three (or four) 'wisps' are seen both in mid-IR and radio on the northern side. The green diamond indicates the radio source, 43.21+61.3 of \citet{rodriguez-rico04}, which coincides with the X-ray source, Xp9, at the position of the radio kinematic centre. The five brightest radio SNRs, all of which have X-ray counterparts, are encircled in white. The radio SNR 41.95+57.5 (=X-ray SNR, Xp15) and presently the brightest radio SNR of SN2008iz \citep{brunthaler09} are indicated.}
\label{fig:jvlaNeII}
\end{figure}

The recent star formation in M82 is believed to be triggered by a tidal interaction with M81 \citep[e,g,][]{yun93}, resulting in two short ($\sim 1$ Myr) bursts of star formation 5 Myr and 10 Myr ago \citep{rieke93,forsterschreiber03}. The first intense burst peaking at the nuclear region was followed by the second burst in a circumnuclear ring. The mid-IR E1 knot is likely clustered H\,{\sc ii} regions and is associated with the eastern part of the starburst ring, just inside the molecular concentration \citep{achtermann95}. After the second burst, which had an initial star formation rate (SFR) of $\sim 43~M_{\sun}$~yr$^{-1}$, the SFR declined rapidly \citep{forsterschreiber03}.
The present SFR, based on the FIR luminosity \citep{kennicutt98}, corrected for the flattened IMF at lower masses \citep{rieke09}, is 4.3 $M_{\sun}$~yr$^{-1}$ over the mass range of 0.1-100 $M_{\sun}$.  

According to the two-burst model, the starburst ring should have accumulated numerous SNRs but also still keeps producing new SNe.
However, some 40 radio SNRs in M82 all avoid the enhanced star-forming knots. While this has been a subject of discussion in various articles  
\citep[e.g.][]{golla96,degrijs01,alonso-herrero03},
it could be understood when taking the starburst age and local SN feedback into account.
We illustrate it using E1 as an example, since it is part of the starburst ring where the second burst took place 5 Myr ago for a duration of 1 Myr.

In the simplified starburst episode assumed above, as a massive star with a mass of $M$ lives for $\sim 30(M/8\,M_{\sun})^{-2.5}$ Myr, stars with masses of $\ga 17M_{\sun}$ already died, leaving both SNe and SNRs. Taking the typical SFR surface density $5\times 10^{-4} M_{\sun}$~ yr$^{-1}$~pc$^{-2}$ for the second burst \citep{forsterschreiber03}, a circular region with a diameter of 3\arcsec\ gives a SFR of $\sim 1~M_{\sun}$~yr$^{-1}$ for E1. Then E1 has accumulated some 2000 SNRs at present time. Here we used the SFR to SN rate conversion factor of 0.0088 of \citet{horiuchi11}.
The series of SN explosions produce a hot cavity and its filling factor within the star-forming knot grows as the number of SNe increases \citep{mckee77}. In this rarefied environment, SNe and SNRs do not cool. This means they directly heat the ambient medium by providing much of the energy in mechanical form and thus could be difficult to detect due to low brightness.

The above process produces metal-rich hot fluid of a temperature of $10^8$ K \citep{chevalier85}. In our {\it Chandra} observations, the region-segment dX34 coincides with the central part of E1 (Fig. \ref{fig:hotEW+neii}). The X-ray spectrum of dX34 shows no Fe\,{\sc xxv} line (Fig. \ref{fig:map_values}), suggesting that the putative hot fluid, too, radiates little with its low emission measure.  
On the other hand, abundant amounts of dust are produced by SNe \citep{gall18}, providing a source of strong FIR emission. The region also acts as a cosmic ray reservoir, since a significant fraction ($\sim 10$\%) of SN explosion energy may go efficiently into cosmic ray acceleration \citep{pais18,grenier15,caprioli14,bell04}.
Thus the local SN feedback and dust production together produce IC X-ray emission, which dominates the dX34 spectrum. 
However, once the metal-enriched hot fluid flows out of E1, it entrains  the surrounding cool ISM. This leads to a drop in temperature,
as demostrated by simulations \citep[e.g.][]{tomisaka88},
and an increased density makes the fluid sufficiently radiative, resulting in sub-$10^8$\,K gas with strong Fe\,{\sc xxv}.
This transition can be illustrated by the spectrum of dX35, which lies immediately outside E1 and at the base of the south-bound Fe\,{\sc xxv} extension III (Fig. \ref{fig:hotEW+neii}). The segments dX34 and dX35 are both located in the disc plane and separated by only $\sim 20$ pc, yet the strong Fe\,{\sc xxv} line in the dX35 spectrum contrasts sharply with that of dX34 with no Fe\,{\sc xxv}.

The above hypothesis explains how, despite lacking detectable SNRs, E1 exhibits the X-ray spectrum dominated by IC emission, yet it acts as a hub of the Fe\,{\sc xxv}-enhanced emission. The total energy deposited by 2000 SNe in E1 over the last 5 Myr amounts to $\sim 2\times 10^{54}$ erg. About 10\% of this energy would go to cosmic rays, and $\sim 5$\% is radiated away via the Fe\,{\sc xxv}-emitting hot gas, as observed. Therefore the majority of the SN energy should have gone to the chimney in knetic form and was transported to the halo.

\subsection{Detectability of radio and X-ray SNRs}

The brightest five radio SNRs have X-ray counterparts \citep[][see Fig. \ref{fig:jvlaNeII}]{iwasawa21}. An inspection of the deep {\it JVLA} image found a faint radio source which appears to coincide with another X-ray SNR, Xp17. This makes six X-ray--radio correspondences out of the seven X-ray SNRs detected in \citet{iwasawa21}, which is remarkably high, compared to those in other nearby galaxies \citep{leonidaki10,long10,pannuti07}, suggesting that they share a favourable condition, for example, dense environment, to be highly radiative in both wavelengths. Such a dense environment no longer exists in the interior of the past burst sites, as described above. Instead, freshly formed giant molecular clouds (GMC), whose lifetime could be a few Myr \citep{dobbs13}, may sustain the present star-forming activity which dispenses young SNRs. In fact, all the seven X-ray SNRs are found to reside within GMCs with masses of $4\times 10^3\,M_{\sun}$ - $1.7\times 10^{7}\,M_{\sun}$, catalogued by \citet{krieger21} who mapped them in CO(1-0) with IRAM NOEMA (the median projected distance of those X-ray SNRs from the respective cloud centres is 34 pc while the median cloud redius 54 pc), supporting the above idea.
They provide an ideal environment for those young radio and X-ray SNRs which are less than a few hundreds yr old \citep[e.g.][]{fenech10,iwasawa21}, since there have been little pre-existing SN feedback. The estimated radio SN rates are in the range of 0.03-0.1\,yr$^{-1}$ \citep{unger84,kronberg85,muxlow94,huang94,fenech08,fenech10}, corresponding to SFR of 3.4-11\,$M_{\sun}$~yr$^{-1}$, in agreement with the current SFR.
Different manifestations of SN feedback depending on their molecular gas environment, as discussed for nearby star-forming galaxies in the PHANGS survey \citep{maykerchen22}, is also in support of our argument.

\subsection{6.4 keV cold Fe K line}

The origin of the 6.4 keV cold Fe K line is unclear.
\citet{liu14} argued irradiation of optically thin \nH $\sim 10^{22}$ \psqcm\ cold ISM by the luminous X-ray binaries can account for the total line luminosity.
The two segments of high surface brightness in 6.4 keV emission (dX14 and 28) near the bright X-ray binary concentration in the western part may fit this scenario.
However, the mapped distribution of the line emission suggests its diffuse nature and the extended emission towards the eastern part is difficult to account for by photoionisation of bright X-ray binaries alone.

Here, we examine an alternative line production mechanism by cosmic ray bombardment, which has been proposed for explaining 6.4 keV Fe lines seen in the Galactic centre \citep{valinia00,dogiel09}
and the surroundings of some of the Galactic SNRs \citep{tatischeff12}.
This process is compatible with the diffuse nature of the emission across the starburst region. While the energy density of cosmic ray in M82 is expected to be high, the line production is inefficient: $\eta = L_{6.4}/L_{\rm CR}\sim 10^{-6}$ at most \citep{tatischeff12}. With the SFR $=33 M_{\sun}$~yr$^{-1}$ over the past 15 Myr, the SN rate would be $\nu_{\rm SN}\sim 0.3$~yr$^{-1}$. Assuming $\epsilon = 10$\% of the SN energy ($10^{51}$ erg per SN) goes to cosmic rays, the efficiency of $\eta = 10^{-6}$ gives an expected 6.4 keV line luminosity of $\sim 1\times 10^{36}(\epsilon / 0.1)(\nu_{\rm SN}/0.3\,{\rm yr}^{-1})(\eta /10^{-6})$\,\ergps, if SNe are the sole accelerator of cosmic rays. This estimate is more than one order of magnitude below the total 6.4 keV line luminosity (Table \ref{tab:feklines}) and cannot be a major source of the detected line emission.
Similar to the Galactic centre \citep[Sgr B2 clouds, in particular, ][]{zhang15, rogers22,kuznetsova22}, while the 6.4 keV line emission of the cosmic ray origin accounts for the low-brightness, extended emission in M82, the brightest spots need to be explained by irradiation of transient flaring sources, whether they are M82 X-1 or transient X-2, which can reach $L_{\rm X} >10^{40}$ \ergps, \citep{kaaret01,brightman19}, or even an active nucleus, as discussed below. A long term monitoring of the brightest regions can test the photoionisation scenario through variability in line flux.

\subsection{Possible AGN at the galactic centre}

The previous claim of a low-luminosity AGN in M82 was based on hard X-ray variability observed with {\it ASCA} \citep{ptak99,matsumoto99}, which, as we know now, is caused by the ULXs. Here, we examine how plausible the faint X-ray source, Xp9, at the galactic centre might be an AGN.
Xp9 has been assumed to be an X-ray binary but the presence of the radio counterpart makes it unique among the X-ray binaries detected in the central region.
The deep {\it JVLA} image verifies the radio counterpart, as marked by a magenta diamond in Fig. \ref{fig:jvlaNeII}, whilst
none of the other X-ray binaries (including the ULXs) have a point-like radio counterpart in the {\it JVLA} image (Xp6 and Xp16 coincide with compact radio sources but they are radio SNRs, \citet{fenech10}). An X-ray binary could have a radio jet in the hard state. However, as Galactic black hole binaries typically have radio luminosity $L_{\rm R}\leq 10^{30}$ \ergps \citep[e.g.][the micro quasar GRS1915+105 occasionally reaches $1\times 10^{32}$ \ergps, as an exception]{gallo12} in the hard state, no radio detection is expected for X-ray binaries in M82, in agreement with the lack of radio counterpart of the other X-ray binaries. In contrast, assuming a flat spectrum and 4 mJy at 5GHz, Xp9 has log $L_{\rm R}\simeq 32.5$ [\ergps], which is comparable to that of Sgr A$^{\star}$. In the context of the black hole fundamental plane \citep{merloni03}, radio luminosity, $L_{\rm R}$, increases with black hole mass, $M_{\rm BH}$, and thus the radio loudness of Xp9 points to a large $M_{\rm BH}$.

The X-ray spectrum of Xp9 is very hard with the power-law slope of $\alpha\sim -1$ and the source is visible only above 3 keV. The hard spectrum is likely due to strong absorption: if an intrinsic slope of $\alpha = 0.7$ is assumed, absorbing column density would be \nH\ $\sim 2\times 10^{23}$ \psqcm. The observed 4-8 keV flux is $3.0\times 10^{-14}$ \ergpspsqcm\ and the absorption-corrected 2-10 keV flux would be $1.2\times 10^{-13}$ \ergpspsqcm. The latter corresponds to log $L_{\rm X}\simeq 38.27$ [\ergps]. The 4-7 keV light curve remains stable over 16 yr of the {\it Chandra} observations \citep{iwasawa21}. The X-ray luminosity is compatible to a bright X-ray binary but could also be of low-luminosity AGN, powered by an intermediate-mass black hole or a supermassive black hole operating at inefficient accretion flow, such as an advection-dominated accretion flow (ADAF) \citep{narayan94}.

\citet{gaffney93} obtained the nuclear enclosed mass $M~(r<8.5\thinspace {\rm pc}) = 3\times 10^7 M_{\odot}$, using the CO absorption bandhead kinematics \citep[see also][]{greco12}. It can be considered as an upper limit of the mass of a putative black hole, $M_{\rm BH}$. However, $M_{\rm BH}\sim 10^6\,M_{\sun}$, similar to that of Sgr A$^{\star}$, would be possible only if a significant portion of the stellar disc was stripped away on the close encounter with M81, leaving M82 as a bulge-dominated galaxy, as proposed by \citet{sofue98}, otherwise it is deemed to be overmassive for the galaxy with the total mass of $10^{10}\,M_{\sun}$ \citep{greco12}. With the observed radio and X-ray luminosities, the black hole fundamental plane gives $M_{\rm BH}$ of $10^3$-$10^4\,M_{\sun}$ \citep{merloni03,glutekin19}. While whether black holes in this mass range exist is presently unclear, it is compatible with the argument of M82 to be a bulgeless galaxy \citep{mayya09}. Some low-luminosity AGNs hosted by small, bulgeless galaxies, such as NGC\,4395, have black hole masses of the order of $10^5\,M_{\sun}$ \citep{peterson05}. When $M_{\rm BH}=10^5\,M_{\sun}$ is assumed, the typical bolometric correction of $\kappa\equiv L_{\rm bol}/L_{\rm X} = 16$ for low-luminosity AGN \citep{ho08} gives the bolometric luminosity of Xp9 to be $3\times 10^{39}(\kappa/16)$ \ergps. Then the Eddington ratio would be $\lambda\sim 2\times 10^{-4}(\kappa /16)(M_{\rm BH}/10^5\,M_{\sun})^{-1}$, well within the ADAF range. The flat radio spectrum is similar to Sgr~A$^{\star}$ and other sources that are suspected to be operating in ADAF, slightly steeper than $\sim 0.3$ predicted for ADAF \citep{mahadevan97}. There is no notable evidence for past elevated AGN activity, such as the {\it Fermi} bubble in our Galaxy, although the starburst activity might disguise it.   

\begin{figure}
\centerline{\includegraphics[width=0.37\textwidth,angle=0]{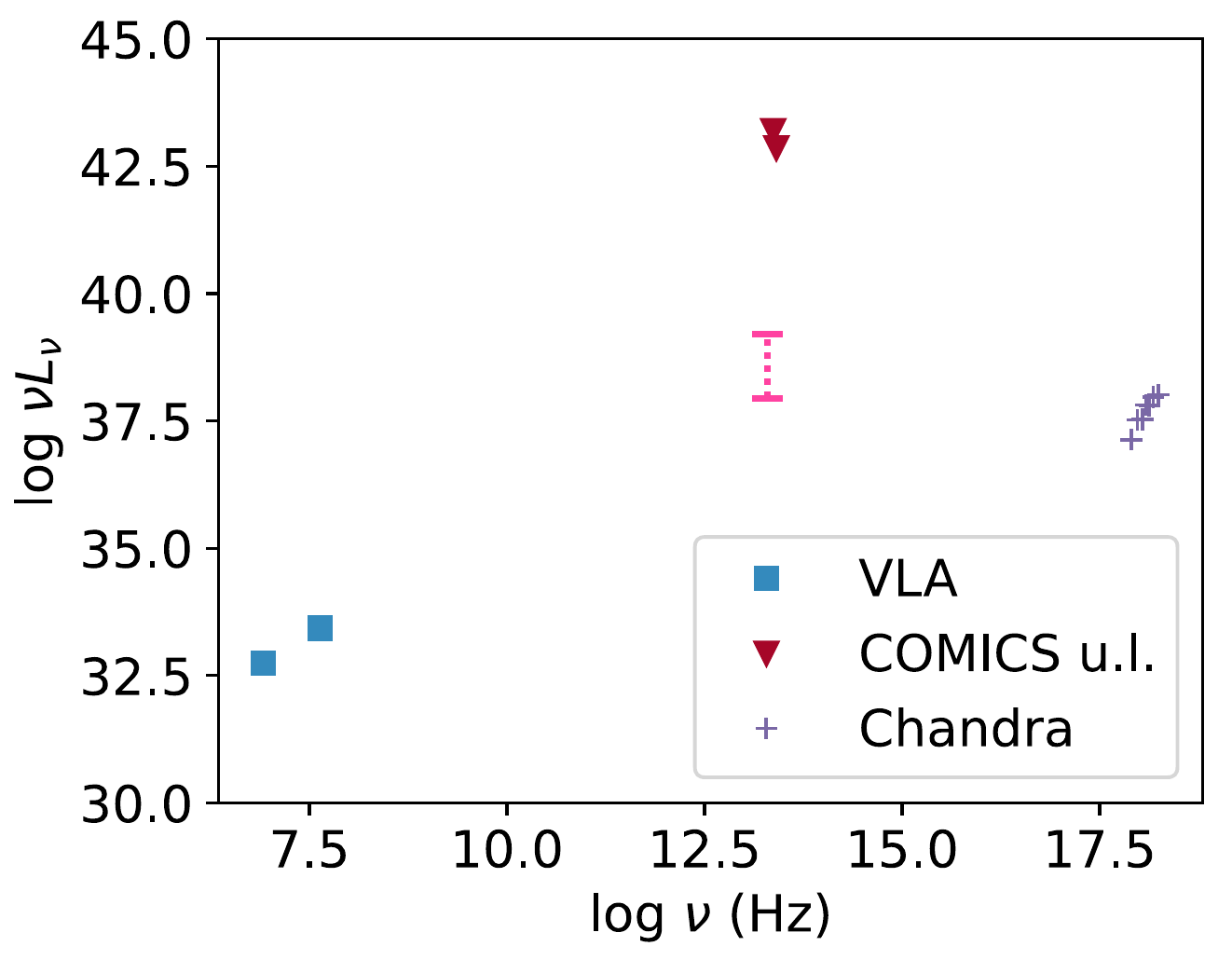}}
\caption{Spectral energy distribution of Xp9. The radio data points are from \citet{rodriguez-rico04}. The mid-IR data points are upper limits obtained from the {\it Subaru}-COMICS images \citep{gandhi11}. The luminosity range at 12\,$\mu $m, expected from the low-luminosity AGN hypothesis, is indicated by the dotted-line, which could be within the reach of the upcoming {\it JWST} observations.}
\label{fig:SEDXp9}
\end{figure}

The radio-to-X-ray spectral energy distribution of Xp9 is shown in Fig. \ref{fig:SEDXp9}. The mid-IR points are upper limits from
\citet{gandhi11}. The 12-$\mu $m luminosity predicted by the ADAF models \citep{mahadevan97,kino00} and the mid-IR--X-ray relation of \citet{gandhi09} is around $10^{38}$-$10^{39}$ \ergps. This is translated to a flux density of 0.7-7 $\mu $Jy, which may be barely detectable with the upcoming  {\it James Webb Space Telescope}-MIRI observation (ID: 1701). As the mid-IR luminosity expected from an X-ray binary is a few orders of magnitude farther down, any detection with {\it JWST} would point to the low-luminosity AGN hypothesis.

\section{Conclusions}

We present the spatially resolved spectral map of the 4-8 keV diffuse emission in M82 at a few arcsec resolution, after removing point-like sources. Our findings are summarised as follows. 

\begin{enumerate}

\item The EW of the Fe\,{\sc xxv} line from hot gas varies widely across the diffuse emission region, including areas with EW $< 0.1$ keV, suggesting the presence of non-thermal emission. The morphology of the diffuse X-ray emission resembles those of the FIR and radio emission, in favour of inverse Compton scattering off the FIR photons by cosmic ray electrons as the origin of the non-thermal emission.

\item We found that the 4-8 keV diffuse emission is composed of three spectral components with distinct origins: 1) inverse Compton emission which carries $\sim 70$\% of the continuum luminosity in the band; 2) the hard tail of the soft X-ray wind emission of $kT\sim 1$ keV; and 3) metal-rich, hot gas emission of $kT\sim 5$ keV, which accounts for the majority of the observed Fe\,{\sc xxv} line but is a minor component of the continuum.

\item The hot gas, traced by enhanced Fe\,{\sc xxv} EW, is found in a limited area near the galactic disc and appears to flow out from the eastern part of the starburst ring and fills the chimneys marked by mid-IR and radio voids. The chimneys dominate in transporting the flow of SN energy from the disc to halo in M82.

\item The brightest, young X-ray and radio SNRs are found to reside in GMCs that are presumably newly formed and thus free from strong SN feedback. This could explain their positional displacement from the past active star-forming sites where any SN would be little radiative due to the hot, low-dendity interior.

\item The faint X-ray source, Xp9, located at the galactic centre, is unusual for an X-ray binary to have a luminous radio counterpart with a flat spectrum, offering a possibility to be a low-luminosity AGN with $L_{\rm X}\sim 10^{38}$ \ergps\ in an ADAF phase. 

\end{enumerate}

A similar technique employed in this work can be applied with advanced future X-ray mission with high angular resolution, such as {\it AXIS} or {\it LYNX}, to reveal effects of feedback on the ISM in a range of nearby galaxies, including those with central AGN.

\begin{acknowledgements}
We thank the referee for helpful comments. This research made use of data obtained from {\it Chandra X-ray Observatory}, of which the archive is maintained by {\it Chandra} X-ray Center (CXC) and NASA/IPAC Extragalactic Databases (NED) and Infrared Science Archive (IRSA), which are funded by the NASA and operated by the California Institute of Technology. Software packages of CIAO~4.12, HEASoft~6.27.1, Astropy, PyMC, Matplotlib, Numpy, Pandas, Scipy and IPython were used for data analysis. KI acknowledges support by grant PID2019-105510GB-C33 funded by MCIN/AEI/10.13039/501100011033 and ``Unit of excellence Mar\'ia de Maeztu 2020-2023'' awarded to ICCUB (CEX2019-000918-M). MPT acknowledges financial support from the Spanish Ministerio de Ciencia e Innovaci\'on (MCIN), the Agencia Estatal de Investigaci\'on (AEI) through the "Center of Excellence Severo Ochoa" award to the Instituto de Astrofísica de Andalucía (SEV-2017-0709) and through grant PID2020-117404GB-C21 funded by MCIN/AEI/10.13039/501100011033.
\end{acknowledgements}

\bibliographystyle{aa} \bibliography{m82diffuse}{}


\begin{appendix}

\section{Fe K lines in thermal emission spectrum}

\begin{figure}
  \centerline{\includegraphics[width=0.5\textwidth,angle=0]{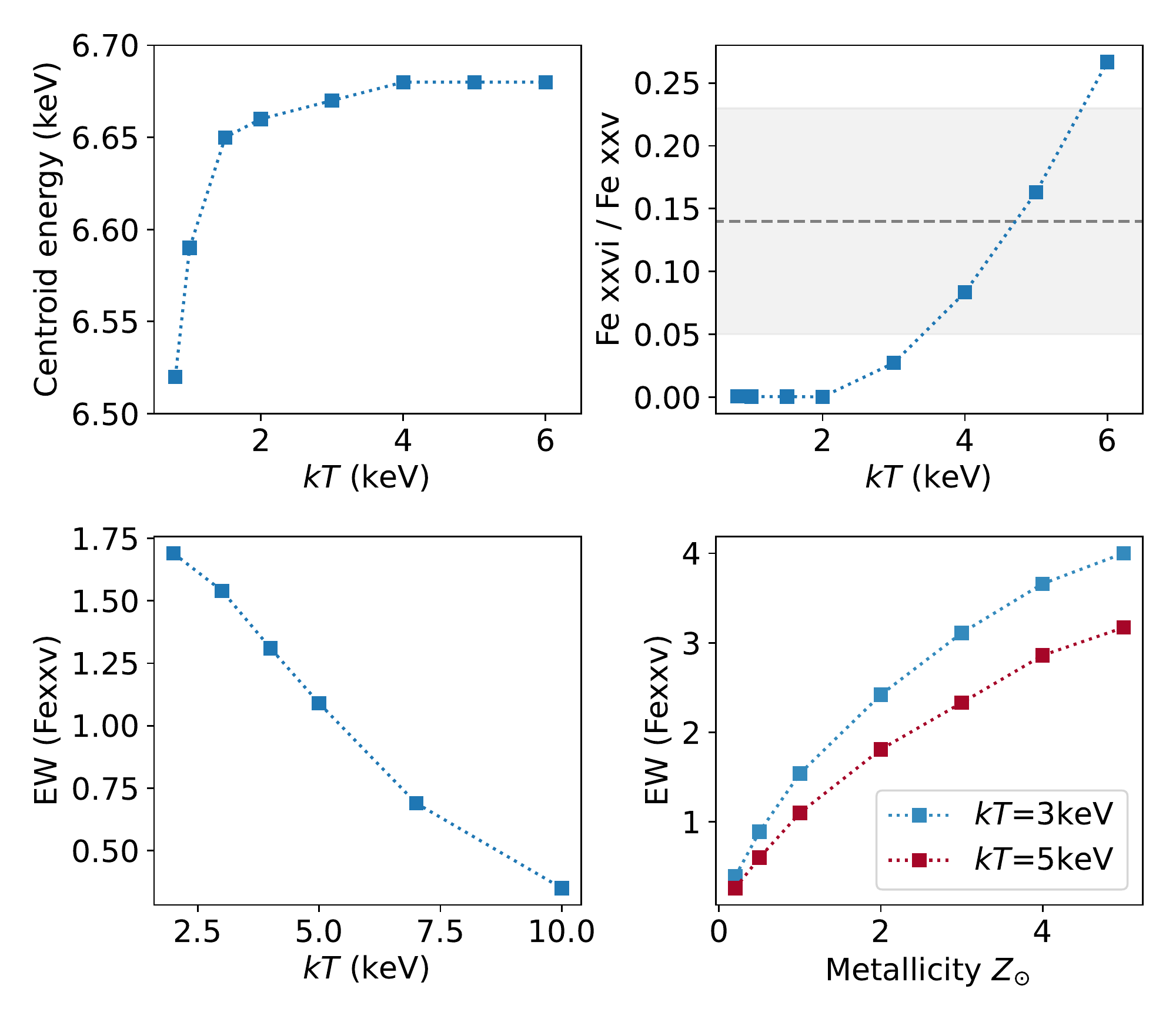}}
  \caption{Fe K line properties of thermal emission spectrum. Top-left: Line centroid energy as a function of gas temperature; Top-right: Line intensity ratio of Fe\,{\sc xxvi} and Fe\,{\sc xxv} as a function of gas temperature. The dashed-line indicates the observed line ratio with 68\% error region in grey shade; Bottom-left: Fe\,{\sc xxv} EW as a function of temperature when the Solar metallity is assumed; and Bottom-right: Fe\,{\sc xxv} EW as a function of metallicity for gas temperature of $kT=3$ keV (blue) and $kT=5$ keV (red).}
    \label{fig:sim}
\end{figure}

Properties of Fe K line emission seen in a thermal emission spectrum, when described by a Gaussian as done in this work, are illustrated in Fig \ref{fig:sim}. We simulated spectra using {\tt apec} assuming the collisional ionisation equilibrium condition and obtained these properties. When the gas temperature $kT$ exceeds 2 keV, Fe\,{\sc xxv} emission becomes prominent. The line centroid energy moves depending on $kT$. When $kT$ is in the range of 4-6 keV, the centroid energy in found at 6.68 keV. Fe\,{\sc xxv} is a triplet which is unresolved at the CCD resolution. A single Gaussian fit would give $\sigma\simeq 30$ eV for the line blend.
The H-like Fe\,{\sc xxvi} emission arises when $kT$ exceeds $\sim $3 keV, and the line intensity ratio between Fe\,{\sc xxvi} and Fe\,{\sc xxv} changes as a function of $kT$. The observed line ratio of $0.14\pm 0.09$ indicates $kT\approx 4.5\pm 1$ keV, if it is a single component gas. Fe\,{\sc xxv} EW varies as a function of temperature at a given metallicity. This is illustrated in the figure in the case of the Solar metallicity. Finally, metallicity dependence of Fe\,{\sc xxv} EW for gas temperatures of $kT=3$ keV and 5 keV is shown. We note that the dependence is not linear but flattened as increasing metallicity.

\section{An unidentified 4.5 keV emission feature}

\begin{figure}  
\centerline{\includegraphics[width=0.37\textwidth,angle=0]{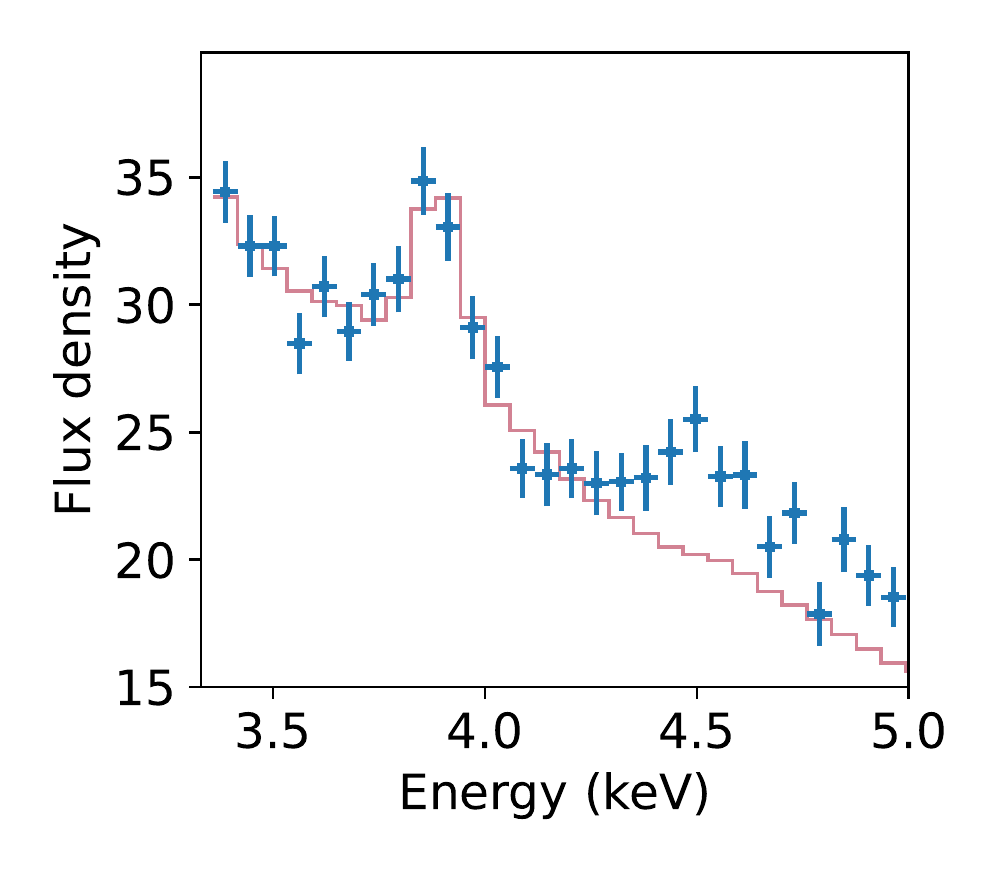}}
  \caption{Diffuse emission spectrum of the central part of M82 in the 3.3-5 keV band. It illustrates the excess emission at 4.5 keV. The red histogram indicates the best fitting {\tt apec} model, which describes the Ca feature at 3.9 keV. No strong atomic line feature that corresponds to the 4.5 keV emission is expected. The flux density is in units of $10^{-14}$ \ergpspsqcmpkeV .}
    \label{fig:e45line}
\end{figure}

There is a possible line-like excess at 4.5 keV in the total diffuse emission spectrum. The Ca K band spectrum at $\sim 4$ keV indicates the temperature $kT\sim 1.8$ keV (Fig. \ref{fig:e45line}). No emission-line feature is expected at 4.5 keV in a thermal emission spectrum of a similar temperature. We examined the 4.1-5 keV data for its detection with a power-law model with or without a narrow Gaussian line. The lower energy bound was chosen to avoid the Ca K emission. To optimise a line detection, we rebinned the data with 60-eV intervals, which samples the major part of a line feature by three bins at 4.5 keV, where the spectral resolution is FWHM $\simeq 140$ eV. The Gaussian centroid is found at $4.51\pm 0.03$ keV. The line intensity is $(7.8\pm 2.3)\times 10^{-7}$ \phpspsqcm, corresponding to a EW of $26\pm 8$ eV. The continuum slope is $\alpha = 1.8\pm 0.3$ when \nH $= 4\times 10^{22}$ \psqcm\ is assumed.
We compared between the fits with and without the Gaussian line, using the Akaike Information Criterion (AIC) and Bayesian Information Criterion (BIC). The relative likelihood of the inclusion of the 4.5 keV line is 20 from BIC and 40 from AIC. The difference comes from the degree of penalisation on extra parameters. The detection of the line feature is reasonably strong. A possible identification is a neutral Ti K at 4.51 keV. The cold Fe K line at 6.4 keV is detected and might share the same origin. However, the 4.5 keV line is far too strong to match the cosmic abundance of Ti, which is $\sim 2$ orders of magnitude smaller than Fe. Although the cosmic rays have much more elevated Ti abundance relative to Fe due to spallation, it is not relevant unless the line production is due to the charge exchange.

\section{Image of dX36 and dX37 region from no transient flaring intervals}


\begin{figure} \centerline{\includegraphics[width=0.48\textwidth,angle=0]{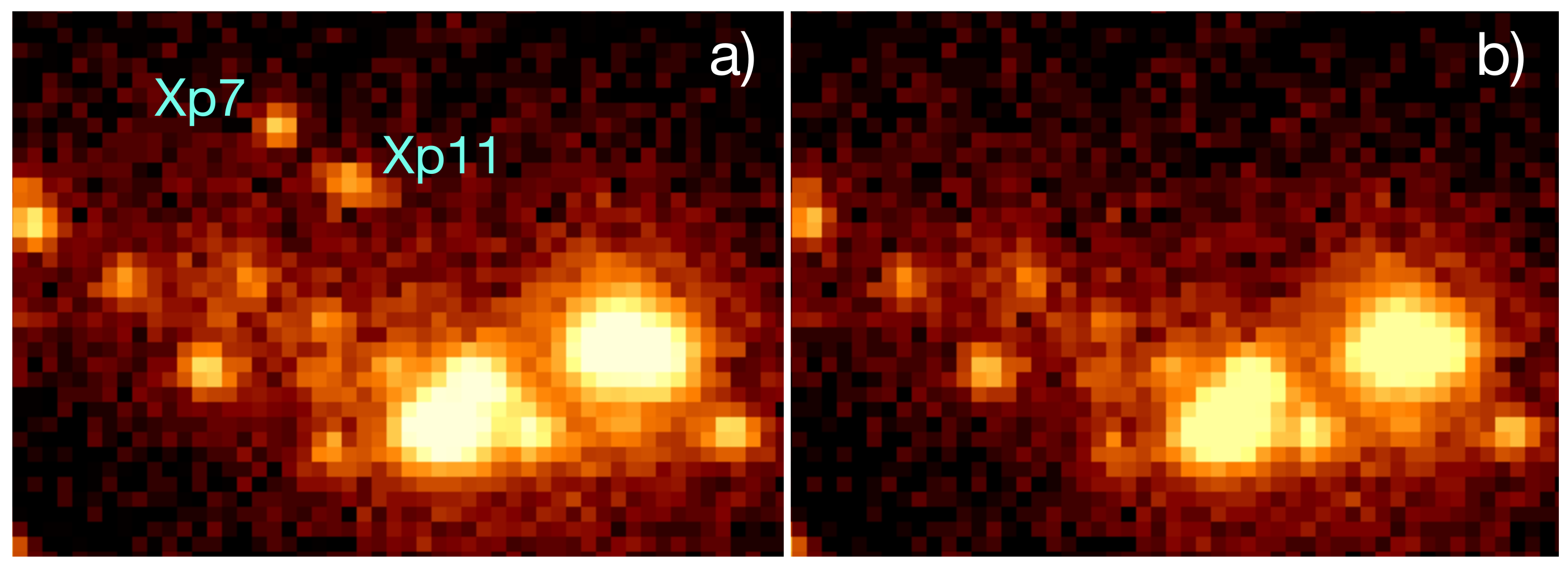}}
\caption{Comparison of the 4-8 keV images of the transient sources Xp7 and Xp11, taken from a) the total 13 exposures; and b) the eight exposures where both sources were in quiescence. The two transient sources are marked on the left panel. }
\label{fig:xp7xp11img}
\end{figure}

The region segments dX36 and dX37 contain the transient sources Xp7 and Xp11, respectively. Xp7 exhibited flares during the first two observations in 1999, while Xp11 during the three observations in 2010 (see the {\it Chandra} observation log in Table 1.). Fig. \ref{fig:xp7xp11img} shows the 4-8 keV image of the region around the two transient sources, obtained from the full 13 exposures (in the left panel) and that of the same region from eight exposures in which neither of them were in the flaring states (in the right panel). No point-like excess is seen at the positions of the two sources in the image when they were in quiescence. This justifies the use of their quiescence data to study the diffuse emission properties in the segments dX36 and dX37.

\section{X-ray extinction map}

\begin{figure}  \centerline{\includegraphics[width=0.45\textwidth,angle=0]{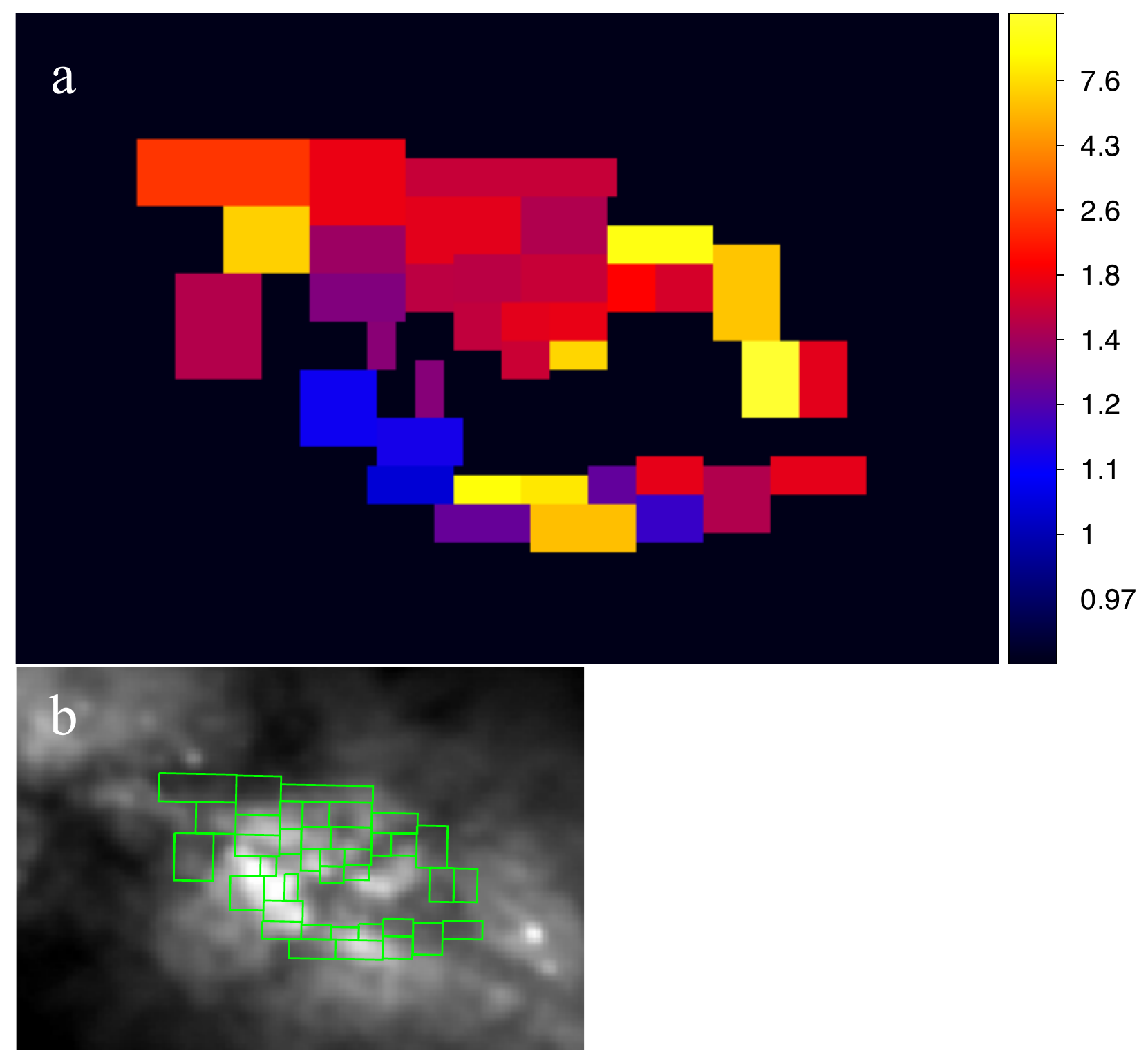}}
  \caption{Extinction across the 4-8 keV diffuse emission region. a) X-ray absorbing column density, \nH, map in units of $10^{22}$ \psqcm. b) The $R$-band image of the central part of M82, taken  by the Bok 2.3-m telescope at Kitt Peak National Observatory, overlaid by the 38 region segments where the X-ray absorbing columns were measured.}
    \label{fig:nH_map}
\end{figure}

The distribution of X-ray absorbing column density over the diffuse emission region is presented in Fig. \ref{fig:nH_map}. The absorbing column density in each region-segment was obtained in two steps. The segment dX31 shows the least absorbed soft-band spectrum and fitting an absorbed {\tt apec} model to the 0.9-1.7 keV spectrum gives a temperature of $kT = 0.87^{+0.02}_{-0.01}$ keV, metallicity of $1.0^{+0.4}_{-0.2} Z_{\odot}$ and \nH\,$=1.11(1.07-1.14)\times 10^{22}$ \psqcm. Firstly, we fitted \nH\ of the spectra of all the other segments in the same band, assuming the same temperature and metallicity. These \nH\ are all in the range of (1-2)\,$\times 10^{22}$ \psqcm. They represent absorption for the soft X-ray wind and might not reflect the absorption towards the inner part where the hard X-ray emission comes from. All the spectra were fitted again but in the 4-7 keV band, assuming $kT=6$ keV. When this hard-band \nH\ exceeds that for the soft band, the hard-band \nH\ was adopted. This applies to dX1, 14, 15, 18, 19, 26, 27, and 28, for which the 4-7 keV continua show a hard slope. It is found that a few regions show \nH\ up to $10^{23}$ \psqcm.

For comparison, the optical image of the region taken in $R$ band by the Bok 2.3-m telescope at Kitt Peak National Observatory, overlaid by the X-ray region segments, is also shown in Fig. \ref{fig:nH_map}. The image was retrieved from NED. The region segments showing high \nH\ values generally coincide with the dust lanes seen in the optical image.


\section{Galactic centre positions}

The positions of Xp9, radio source 43.21+61.3, the three kinematic centre measurements, the nuclear position measured by WISE ($W1$) in the WISE eXtended Source Catalogue (WXSC), and the two 2.2\,$\mu $m peak measurements, plotted in Fig \ref{fig:galcentres}, are reproduced in the Table below.

\begin{table}
  \caption{Galactic centre positions}
  \label{tab:galcentre}
  \begin{tabular}{lccc}
\hline\hline
& R.A.(J2000) & Dec. (J2000) & ref\\
\hline
Xp9 & 52.0 & 47.3 & 1 \\
43.21+61.3 & 51.95 & 47.2 & 2 \\
H~{\sc i} kinematic centre & 51.9 & 47.2 & 3 \\
Ne~{\sc ii} kinematic centre & 52.14 & 46.4 & 4 \\
H92$\alpha $ ring centre & 52.0 & 46.9 & 5\\
WISE & 51.95 & 47.2 & 6 \\[5pt]
$2.2\mu $m peak (R80) & 52.6 & 47.2 & 7 \\
$2.2\mu $m peak (L90) & 52.3 & 46.2 & 8 \\
\hline
\end{tabular}
\tablefoot{Positions in J2000. The RA and Dec are given as the offset from RA = $09^{\rm h}55^{\rm m}00^{\rm s}$, Dec = $+69\degr 40\arcmin 00\arcsec$. 1: X-ray source in \citet{iwasawa21}; 2: Radio source in \citet{rodriguez-rico04}; 3: Radio kinematic centre of \citet{weliachew84}; 4: Ne\,{\sc ii} kinematic centre with an ellipse ring fit of \citet{achtermann95}; 5: H92$\alpha $ ring centre from \citet{rodriguez-rico04}; 6: WISE nuclear position of \citet{jarrett19}; 7: 2.2\,$\mu $m peak from \citet{rieke80}; and 8: \citet{lester90}. The positional errors are all $\sim 0.5\arcsec $ except for 1.5\arcsec\ for the declination of the $2.2\mu$m peak of \citet{rieke80}.}
\end{table}

\section{X-ray spectra of the diffuse emission from the 38 segments}

\begin{figure*}
  \centerline{\includegraphics[width=0.99\textwidth,angle=0]{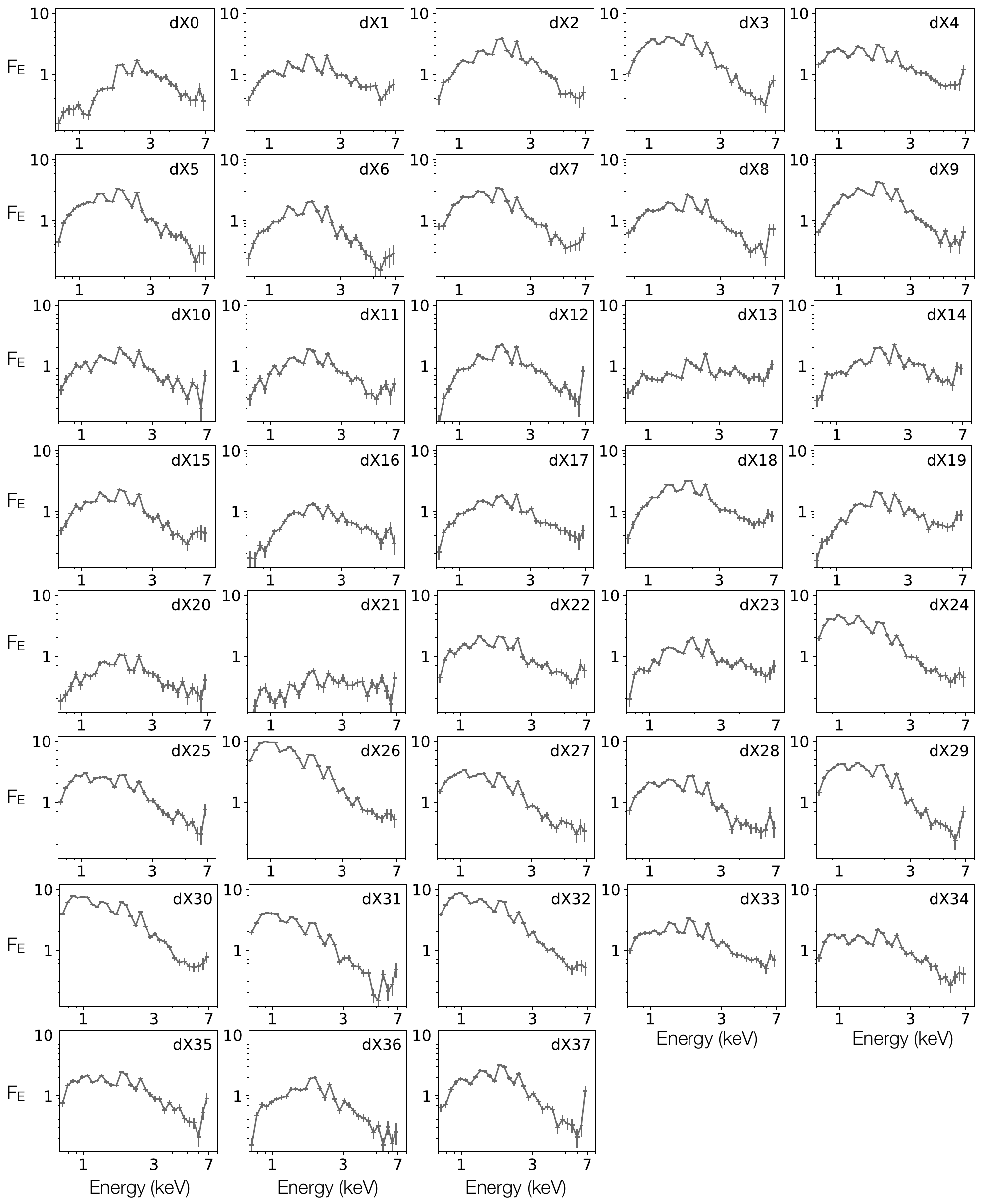}}
\caption{0.7-7 keV {\it Chandra} spectra of the 38 region segments, dX0 through dX37, over the diffuse emission. The flux density, $F_{\rm E}$, is in units of $10^{-14}$ erg s$^{-1}$ cm$^{-2}$ keV$^{-1}$. The two highest energy bins corresponds to the two Fe-line bands of cold Fe K at 6.4 keV and Fe\,{\sc xxv} at 6.7 keV. }
\label{fig:dxbbsp}
\end{figure*}

\end{appendix}

\end{document}